\documentclass[a4paper,conference,10pt]{IEEEtran}
\IEEEoverridecommandlockouts
\usepackage{graphicx,graphics,epsfig,epstopdf,float}
\usepackage{amsmath} 
\usepackage{amssymb}  
\usepackage{mathrsfs}
\usepackage{enumerate}
\usepackage{subfigure}
\usepackage{bm}
\usepackage{psfrag}
\usepackage{cite}
\usepackage{mdwlist}
\usepackage{longtable}
\usepackage{array}
\usepackage{arydshln}
\usepackage{acronym}  
\usepackage{verbatim}

\usepackage{diagbox}
\usepackage{bm}
\usepackage{multirow}
\usepackage{tikz}
\usetikzlibrary{shapes.geometric, arrows.meta, positioning, calc, fit, backgrounds, decorations.pathmorphing}
\usepackage{flushend}
\providecommand{\msf}[1]{\mathsf{#1}}  
\providecommand{\mcal}[1]{\mathcal{#1}} 
\newcommand{\rv}[1]{\MakeLowercase{\msf{#1}}} 
\newcommand{\RV}[1]{\bm{\MakeLowercase{\msf{#1}}}}  
\newcommand{\RS}[1]{\MakeUppercase{\msf{#1}}} 

\newcommand{\V}[1]{\bm{#1}} 
\newcommand{\M}[1]{\bm{#1}} 
\newcommand{\Set}[1]{\mcal{#1}} 

\usepackage{booktabs}
\usepackage{cite}
\usepackage{amsmath,amssymb,amsfonts}
\usepackage{algorithm}
\usepackage{algpseudocode}
\usepackage{graphicx}
\usepackage{textcomp}
\usepackage{xcolor}
\usepackage{extarrows}

\usepackage[top=1.85cm, bottom=4.35cm, left=1.41cm, right=1.41cm]{geometry}

\def\BibTeX{{\rm B\kern-.05em{\sc i\kern-.025em b}\kern-.08em
    T\kern-.1667em\lower.7ex\hbox{E}\kern-.125emX}}
\begin{document}
\bstctlcite{IEEEexample:BSTcontrol}
\title{Spatiotemporal Tracking in Cooperative ISAC Networks: A Stochastic Geometry Framework\\
}

\author{
	\vspace{0.2cm}
 Bowen~Wang,
 Nanxi~Li,
 Jingzhou~Wu,
 Zheng~Jiang,
 Jianchi~Zhu
\thanks{This research was supported by the National Key Research and Development Program of China under Grant 2024YFE0200102.}
\thanks{B. Wang, N. Li, J. Wu, Z. Jiang, J. Zhu are with the Wireless Technology Research Institute, China Telecom Research Institute, Beijing, 102209, China. (e-mail: \{wangbw2, linanxi, wujingzhou, jiangzheng, zhujc\}@chinatelecom.cn.
).
}
}

\maketitle

\begin{abstract}
We adopt a stochastic-geometry framework to study continuous target tracking in integrated sensing and communication (ISAC) networks, with base-station locations modelled as a Poisson point process. The single-BS analysis shows that the antenna energy-conservation identity forces the mean inter-BS coupling gain to unity, making densification an antenna-irreducible liability for monostatic sensing, while a first-passage-time analysis reveals a target-distance-dependent beamwidth trap. These findings rule out single-BS tracking under densification, motivating a multi-BS cooperative treatment. The static-cluster cooperative mean tracking lifetime is then shown to exhibit a sharp percolation phase transition, with the resulting sensing-capacity ceiling saturating above a critical macro density. Yet the static-cluster idealisation itself misrepresents modern network deployments, where the cooperating cluster is dynamically re-selected as the target drifts; we therefore lift this assumption with a dynamic clustering model that maps the $K$-nearest-neighbour handover onto a 2D Brownian motion with stochastic resetting, and obtain a Bessel-function closed form for the dynamic mean tracking lifetime that dissolves the phase transition under any positive handover rate. With a per-link reliability floor, the dynamic clustering framework preserves classical linear density scaling throughout the realistic 6G regime and delivers an order-of-magnitude capacity lift at small-cell densities. Monte-Carlo simulations corroborate all theoretical predictions.
\end{abstract}

\begin{IEEEkeywords}
stochastic geometry, target tracking, Integrated sensing and communication (ISAC), Poisson point process, dynamic clustering, Brownian motion with stochastic resetting.
\end{IEEEkeywords}

\section{Introduction}
\label{sec:intro}
Continuous tracking of mobile targets such as low-altitude UAVs and connected vehicles is increasingly regarded as a defining capability of next-generation cellular networks \cite{liu2022survey, cui2021integrating}. Network-level analyses of integrated sensing and communication (ISAC), however, have so far focused mainly on static target snapshots and coverage probabilities \cite{zheng2019radar, hua2024performance}, leaving the question of how the network actually behaves while a target is in motion largely open \cite{yuan2023spatio}.

Network-level performance analysis in this space has been built on stochastic geometry \cite{andrews2011tractable, haenggi2012stochastic}, with base-station locations modelled as a Poisson point process and signal-to-interference ratios characterised through their Laplace functionals. This toolkit, originally developed for communication networks, has been adapted to ISAC by accounting for the two-way $\mathcal{O}(r^{-4})$ radar attenuation, the BS-to-BS cross-link interference (CLI) under the asynchronous monostatic working mode, and the 3GPP TR~38.901 LoS-blockage thinning between base stations \cite{bai2014blockage}. On the cooperative side, $K$-nearest-neighbour (NN) clustering combined with Boolean coverage processes \cite{wang2022cooperative, bjornson2020scalable, meester1996continuum} has been used to characterise the geometric gain of aggregating non-coherent echoes from multiple base stations.

Two recurring idealisations, however, limit the predictive power of these analyses. First, the target is typically assumed static once placed, so the analysis returns an instantaneous coverage probability rather than the time the network can actually hold the target in view. Real low-altitude UAVs and connected vehicles exhibit strong spatiotemporal randomness, naturally modelled as continuous-time Brownian diffusion of intensity $D$; capturing how long a target stays within its serving region therefore calls for a first-passage-time (FPT) perspective \cite{redner2001guide}. Second, multi-BS cooperative analyses commit the cooperating $K$-NN cluster once and hold it fixed throughout the tracking session, an assumption that does not match real cloud-radio-access-network (C-RAN) deployments in which the cluster is dynamically re-selected as the target drifts.

This paper develops a unified stochastic-geometry framework that addresses both idealisations. The target is modelled as 2D Brownian motion with diffusion coefficient $D$, and tracking failure is treated as the FPT of the target out of its serving region, so that the metric is a tracking lifetime rather than a coverage snapshot. The cooperating $K$-NN cluster is allowed to be re-selected at Poisson-rate epochs, an operation that maps --- after a translation of coordinates --- onto Brownian motion with stochastic resetting \cite{evans2011diffusion, evans2014resettingsearch, evans2020stochastic}. This dynamic-cluster formulation is consistent with the operational logic of measurement-event-triggered handover in C-RAN and contains the static-cluster baseline as its zero-handover special case, so the static-vs-dynamic comparison can be drawn within a single framework. The asynchronous-monostatic CLI and clutter terms are absorbed into the stochastic-geometry interference functional, and the network kinematics is closed off by an $M/M/\infty$ traffic queue on the target arrival/sojourn process.

The main contributions are:
\begin{itemize}
\item {Single-BS densification collapse with $\bar G\!\approx\!1$.} The antenna energy-conservation identity is shown to pin the mean BS-to-BS coupling gain at unity, identifying densification collapse as antenna-irreducible; Theorem~2 further establishes the beamwidth trap $\theta_m\!\le\!\theta_{\mathrm{crit}}(r_0)$ as a strict physical-layer feasibility constraint.

\item {Static percolation phase transition and heavy-tail MTLT (baseline).} Theorem~1 proves a sharp sub-/super-critical phase transition of cooperative MTLT; Corollary~4 establishes a universal $\tau^{-1.05}$ tail at criticality. These form the $\nu_h\!=\!0$ baseline of the dynamic framework below.

\item {Dynamic-cluster MFPT closed form.} Mapping $K$-NN handover onto Brownian motion with stochastic resetting \cite{evans2011diffusion,evans2014resettingsearch,evans2020stochastic} yields the exact closed form $\mathbb{E}[\rv{T}_{\mathrm{loss}}^{\mathrm{dyn}}]\!=\!\nu_h^{-1}[I_0(R\sqrt{\nu_h/D})-1]$, with the static phase transition recovered as the $\nu_h\!\to\!0$ limit and dissolved under any positive handover rate.

\item {Doubly-logarithmic handover cost and $K^{*}\!=\!K_{\mathrm{rel}}$ collapse.} Inversion of the closed form gives a minimum handover rate doubly logarithmic in the static optimum $K^{*}_{\mathrm{static}}(\lambda_b)/K$, and the operational dynamic cluster size collapses to the per-link availability floor $K_{\mathrm{rel}}\!\in\!\{2,3,4\}$ for all densities, while the static $K^{*}_{\mathrm{static}}(\lambda_b)$ continues to inflate linearly with $\lambda_b$.

\item {Lifted capacity ceiling.} The dynamic capacity ceiling retains classical $\rho_{\max}\lambda_b/(K_{\mathrm{rel}}\rho_0)$ scaling throughout realistic 6G densities, whereas the static ceiling plateaus at $\sim\!265\,\mathrm{targets/km}^{2}$ above $\lambda_b\!\approx\!80\,\mathrm{BS/km}^{2}$, opening a $\sim\!12\times$ capacity gap and $\sim\!13\times$ per-target cluster-commitment reduction at the realistic small-cell ceiling $\lambda_b\!=\!10^{3}\,\mathrm{BS/km}^{2}$.
\end{itemize}

\begin{figure}[!t]
\centering
\includegraphics[width=\columnwidth]{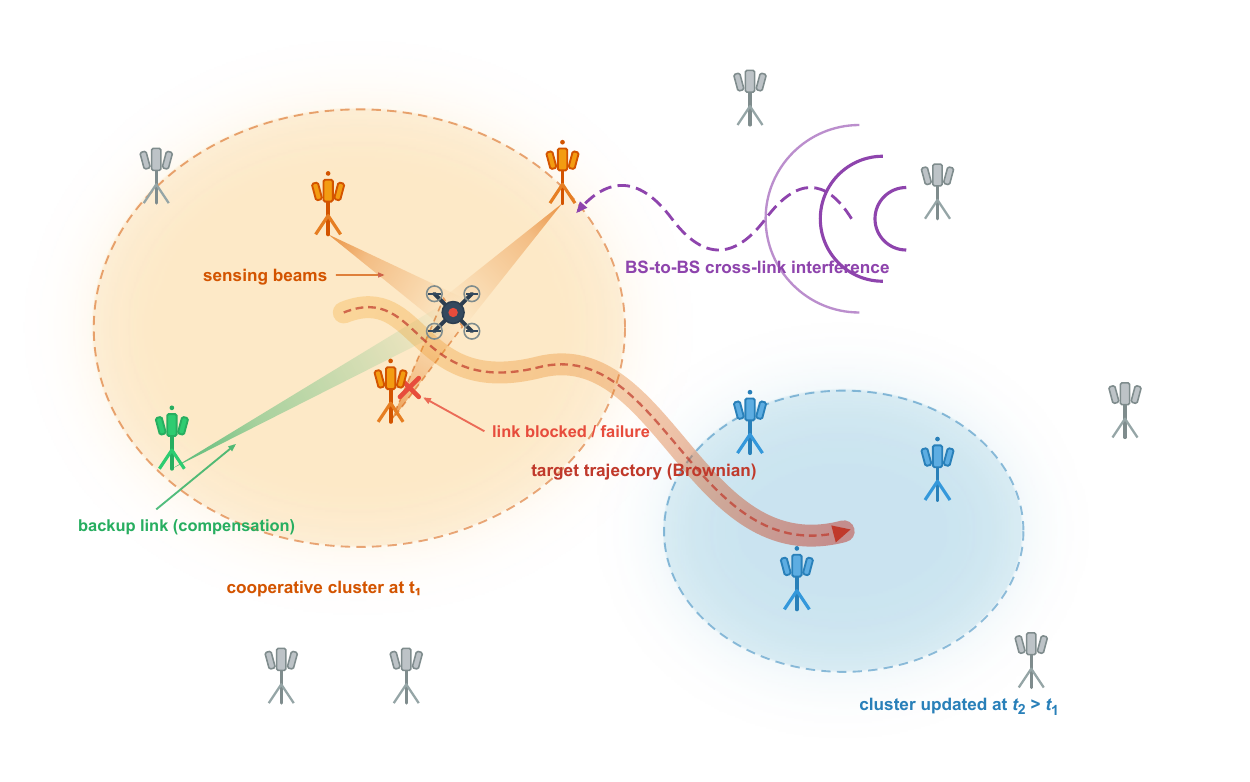}
\caption{Cooperative ISAC target tracking with dynamic K-NN cluster re-selection: a network of base stations cooperatively tracks a moving target; a $K$-NN cluster (orange shaded region) commits sensing resources and combines echoes while suffering cross-link interference from non-cluster BSs. As the target drifts, the cluster is dynamically re-selected to its new vicinity.}
\label{fig:scenario}
\end{figure}

\section{System Model}\label{sec2}
\subsection{Working Mode and Scope}\label{sec:mode}
We focus on asynchronous monostatic ISAC: each BS executes its own transmit/receive cycle without network-wide TDD synchronisation, so when $\text{BS}_0$ is in its receive window every other BS is, with probability one, simultaneously transmitting. This worst-case interference assumption renders all results conservative bounds. The user-centric cooperative cluster $\Set{C}_K$ introduced in Section~\ref{sec:cooperative_tracking} performs cooperative monitoring with non-coherent envelope-fusion at the CU (no carrier-phase, PRF, or coherent multi-static processing). Because BS emissions at $\text{BS}_0$'s receiver are electromagnetically indistinguishable regardless of functional label (downlink data, sensing waveform, or ISAC dual-function), we aggregate all neighbouring BS emissions into a single CLI term $I_{\text{dir}}$; bistatic target-reflected energy is 40--60~dB below the direct path \cite{bai2014blockage} and is absorbed into the clutter aggregate $C_{\text{clutter}}$.

\subsection{Network Topology and Directional Beamforming}
BS locations form a 2D HPPP $\RS{\Phi}_b\!\subset\!\mathbb{R}^2$ of density $\lambda_b$ with hard-core minimum spacing $d_{\min}^{(b)}$ (planning-induced, tens of metres for macro tiers). The hard-core constraint is essential: without it $\mathbb{E}[I_{\text{dir}}]$ diverges at the free-space exponent $\alpha\!=\!2$. We adopt the standard stochastic-geometry approximation of replacing the Mat\'ern-I process by an HPPP of density $\lambda_b$ with a single $\Set{B}(0,d_{\min}^{(b)})$ exclusion at the typical receiver \cite{haenggi2012stochastic}, preserving all $\lambda_b$-linear scalings.

Each BS uses a sectored beam: $G(\theta)\!=\!G_m$ if $|\theta|\!\le\!\theta_m/2$ and $G(\theta)\!=\!G_s\!=\!\zeta G_m$ otherwise ($\zeta\!\in\![0,1)$ the side-lobe nulling factor). Energy conservation forces
\begin{equation}
    G_m = \frac{2\pi}{\theta_m + \zeta(2\pi - \theta_m)} \approx \frac{2\pi}{\theta_m},
    \label{eq:antenna_gain_conservation}
\end{equation}
the inverse coupling $G_m\propto\theta_m^{-1}$ being the fundamental antenna bottleneck. Let $\rho_{\max}\!\in\!(0,1]$ denote the per-BS upper bound on time-frequency fraction allocated to sensing; the unused portion is available for downlink communication.

\subsection{Target Spatiotemporal Dynamics}
Unlike conventional static communication users, sensing targets exhibit pronounced spatiotemporal randomness. We formulate the target dynamics as a stochastic system driven by a spatial birth-death process and continuous-time Brownian motion.

\subsubsection{Poisson Birth-Death Process}
The arrival of targets into the sensing region is modeled as a spatial homogeneous Poisson process with an arrival rate $\lambda_a$. The duration of time a target resides in the region (i.e., the sojourn time) follows an exponential distribution with mean $1/\mu_{\text{sj}}$ (the subscript ``sj'' anticipates the Lagrange-multiplier notation $\nu_1,\nu_2$ of Sec.~\ref{sec:joint_optimization} and disambiguates the sojourn-rate symbol). According to spatial queueing theory (specifically, the $M/M/\infty$ queue model), the total number of active targets present in the network at any arbitrary time $t$ follows a Poisson distribution with parameter $\lambda_a/\mu_{\text{sj}}$.

\subsubsection{Brownian Motion Model}
For any actively tracked target, $\RV{X}(t)=\RV{X}(0)+\sqrt{2D}\,\RV{W}(t)$, where $\RV{W}(t)\sim\mathcal{N}(\V{0},t\M{I}_2)$ is the 2D standard Brownian motion; the mean-squared displacement is $\mathbb{E}\|\RV{X}(t)-\RV{X}(0)\|^{2}=4Dt$. The diffusion $D$ ranges over $0.01$--$0.1\,\mathrm{m^2/s}$ (pedestrians), $0.5$--$2\,\mathrm{m^2/s}$ (urban UAVs), and $5$--$20\,\mathrm{m^2/s}$ (fast vehicles), and is treated as an abstract kinematic descriptor throughout.

\subsection{ISAC Physical Layer and SINCR Formulation}
Consider a typical serving BS, denoted as $\text{BS}_0$, located at the origin $\RV{x}_0 = (0,0)$, which is tracking a target at an instantaneous distance $r_0(t) = \|\RV{X}(t)\|$. Based on the fundamental radar equation and the realistic physical mechanism of network interference, the Signal-to-Interference-plus-Clutter-plus-Noise Ratio (SINCR) at the receiver of $\text{BS}_0$ is defined as:
\begin{equation}
    \text{SINCR}(t) = \frac{S_{\text{target}}(t)}{I_{\text{dir}} + C_{\text{clutter}} + W_0}
    \label{eq:sincr}
\end{equation}
where $W_0$ is the additive white Gaussian noise (AWGN) power. The remaining random variables mapping to the stochastic geometry framework are defined as follows:

\subsubsection{Effective Target Echo ($S_{\text{target}}$)}
The useful sensing signal experiences round-trip (double-path) path loss and target radar cross-section (RCS) fluctuations. The received echo power is given by:
\begin{equation}
    S_{\text{target}}(t) = P_t G_m^2 \sigma \rv{h}_0 r_0(t)^{-4}
\end{equation}
where $P_t$ is the BS transmit power, $\sigma$ is the target RCS, and $\rv{h}_0 \sim \exp(1)$ represents the round-trip small-scale fading, corresponding to the Rayleigh fading assumption (equivalent to the Swerling I/II target fluctuation models).

\subsubsection{BS-to-BS Direct-Path Interference ($I_{\text{dir}}$)}
This is the predominant performance bottleneck. Following 3GPP TR~38.901 \cite{bai2014blockage}, the link between $\text{BS}_i$ and $\text{BS}_0$ is in LoS with probability $p_{\text{LoS}}(v)=e^{-\beta v}$, $v=\|\RV{x}_i-\RV{x}_0\|$, where $1/\beta$ is the LoS coherence length; NLoS interferers are negligible. Under this thinning the residual LoS interferers form an inhomogeneous PPP $\tilde{\RS{\Phi}}_b$ of intensity $\lambda_b\,p_{\text{LoS}}(v)$, propagating with $v^{-2}$ free-space attenuation:
\begin{equation}
    I_{\text{dir}} = \sum_{\RV{x}_i \in \tilde{\RS{\Phi}}_b \setminus \{\RV{x}_0\}} P_t\, G_{tx}(\theta_{i0})\, G_{rx}(\phi_{i0})\, |\rv{g}_i|^2\, \|\RV{x}_i\|^{-2},
\end{equation}
with Rayleigh fading $|\rv{g}_i|^2\!\sim\!\exp(1)$ and $\theta_{i0},\phi_{i0}$ the relative transmit/receive angles between $\mathrm{BS}_i$ and $\mathrm{BS}_0$ (uniform on $[0,2\pi)$ by the stationarity of $\RS{\Phi}_b$). The hard-core spacing $d_{\min}^{(b)}$ enforced above ensures that mean-based quantities such as $\mathbb{E}[I_{\text{dir}}]\!=\!2\pi\lambda_b P_t\bar{G} E_{1}(\beta d_{\min}^{(b)})$ (with $E_1$ the exponential integral and $\bar G$ derived in \eqref{eq:Gbar_explicit}) remain finite and scale as $\Theta(\lambda_b)$.
\label{eq:mean_Idir}

\subsubsection{Environmental Clutter Field ($C_{\text{clutter}}$)}
We model the environmental scatterers (e.g., buildings and foliage) as an independent HPPP $\RS{\Phi}_c$ with density $\lambda_c$. Both self-clutter and cross-clutter are aggregated as the integral of scattered energy over the clutter field:
\begin{equation}
    C_{\text{clutter}} = \sum_{y_k \in \RS{\Phi}_c} P_t G_c \sigma_c |\rv{h}_k|^2 \|y_k\|^{-\alpha_c}
\end{equation}
where $G_c$ is the equivalent beam gain towards the scatterer $y_k$, $\sigma_c$ is the scattering coefficient, $|\rv{h}_k|^2$ is the fading gain, and $\alpha_c$ is the clutter path-loss exponent.

\subsection{Modeling Assumptions and Conservative-Bound Stance}\label{sec:modeling_assumptions}
The framework makes a small set of explicitly conservative modelling choices, listed below. All closed-form theorems can thus be read as performance lower bounds on realistic ISAC deployments.
\begin{itemize}\itemsep0pt
\item[\textit{(A1)}] Asynchronous monostatic working mode (maximal BS-to-BS CLI).
\item[\textit{(A2)}] Brownian target dynamics (no Kalman/particle-filter prediction).
\item[\textit{(A3)}] Poisson exponential-resetting handover (real 3GPP NR is event-triggered and proximity-correlated).
\item[\textit{(A4)}] $M/M/\infty$ target traffic (no bursty arrivals or heavy-tailed sojourns).
\item[\textit{(A5)}] Per-BS sensing budget $\rho_{\max}$ as monolithic resource aggregating RF, CU compute, and fronthaul.
\item[\textit{(A6)}] Per-cooperator per-target resource $\rho_0$ with no beam reuse across co-directional targets.
\end{itemize}
Each loosens, rather than tightens, the bounds below, so the engineering conclusions translate \textit{a fortiori} to real deployments.

\section{Single-BS Tracking Vulnerability Analysis}
\label{sec:single_bs_tracking}

This section analyses single-BS sensing under interference. We first derive the instantaneous coverage probability via the Laplace transform of the interference field, then model the tracking outage as an FPT problem for Brownian motion, leading to the densification paradox.

\subsection{Instantaneous Sensing Coverage Probability}
We first evaluate the static detection reliability. For a target located at a distance $r_0$ and an angle $\theta_0$ within the main lobe ($|\theta_0| \le \theta_m/2$), the instantaneous sensing coverage probability $P_{\text{cov}}(r_0)$ is defined as the probability that the received SINCR exceeds a predefined radar detection threshold $\gamma_s$:
\begin{equation}
    P_{\text{cov}}(r_0) = \mathbb{P}\left(\text{SINCR}(r_0) > \gamma_s \right).
\end{equation}

Recalling the SINCR expression in \eqref{eq:sincr} and the Rayleigh fading assumption for the round-trip channel ($\rv{h}_0 \sim \exp(1)$), we can explicitly extract $\rv{h}_0$ to rewrite the coverage probability using the properties of the exponential distribution:
\begin{align}
    P_{\text{cov}}(r_0) &= \mathbb{P}\left( \rv{h}_0 > \frac{\gamma_s r_0^4}{P_t G_m^2 \sigma} (I_{\text{dir}} + C_{\text{clutter}} + W_0) \right) \nonumber \\
    &= e^{-s W_0} \mathcal{L}_{I_{\text{dir}}}(s) \mathcal{L}_{C_{\text{clutter}}}(s)
    \label{eq:coverage_prob}
\end{align}
where $s = \frac{\gamma_s r_0^4}{P_t G_m^2 \sigma}$, and $\mathcal{L}_{I_{\text{dir}}}(s)$ and $\mathcal{L}_{C_{\text{clutter}}}(s)$ are the Laplace transforms of the direct-path interference and the environmental clutter, respectively.

\textit{BS-to-BS coupling gain.} Each BS points its main lobe at its own committed target; relative to an arbitrary neighbour the main-lobe direction is uniformly distributed on $[0,2\pi)$, so $\text{BS}_i$ irradiates $\text{BS}_0$ with the main lobe with probability $p\!\triangleq\!\theta_m/(2\pi)$ and with the side lobe otherwise. The transmit gain is therefore a discrete random variable with $\mathbb{E}[G_{tx}]=p\,G_m+(1-p)\,G_s$, and by independence of the two BSs' orientations the mean coupled gain is
\begin{equation}
    \bar{G}\;\triangleq\;\mathbb{E}\!\left[G_{tx}\,G_{rx}\right] \;=\; \big(p\,G_m + (1-p)\,G_s\big)^{2}.
    \label{eq:Gbar_explicit}
\end{equation}
Substituting $G_s=\zeta G_m$ and invoking energy conservation \eqref{eq:antenna_gain_conservation} in the form $G_m\,[p+(1-p)\zeta]=1$ gives $\mathbb{E}[G_{tx}]=1$ exactly under the sectored-beam idealisation, so
\begin{equation}
    \bar{G} \;=\; 1\qquad\text{(exact, by energy conservation).}
    \label{eq:Gbar_approx}
\end{equation}
Although $G_m=30\,\mathrm{dBi}$, $\bar G=1$ because the $p^2$ probability of main-on-main coincidence and the $1/p^2$ gain amplitude cancel out; side-lobe nulling ($\zeta\to 0$) does not reduce $\bar G$ below unity. The BS-to-BS interference exponent in the Laplace functional is therefore $\Theta(\lambda_b)$ with a unit coefficient, so densification cannot be offset by antenna sharpening alone.

\textbf{Lemma 1 (Laplace Transform of BS-to-BS CLI):} \textit{Under the blockage-thinned HPPP with $p_{\text{LoS}}(v)\!=\!e^{-\beta v}$, hard-core spacing $d_{\min}^{(b)}$, and sectored beam \eqref{eq:antenna_gain_conservation}, applying the PGFL of inhomogeneous PPPs and Rayleigh identity $\mathbb{E}_g[e^{-A|g|^2}]\!=\!(1\!+\!A)^{-1}$ yields}
\begin{equation}
    \mathcal{L}_{I_{\text{dir}}}(s) = \exp\!\left( -2\pi \lambda_b\, \mathbb{E}_{\tilde{G}}\!\!\int_{d_{\min}^{(b)}}^{\infty}\!\! \frac{s P_t \tilde{G}\, e^{-\beta v}}{v^{2}+s P_t \tilde{G}}\, v\, dv \right),
    \label{eq:lemma1_main}
\end{equation}
\textit{where $\tilde{G}\!=\!G_{tx}(\theta) G_{rx}(\phi)$. The blockage kernel $e^{-\beta v}$ kills the large-$v$ divergence at $\alpha\!=\!2$ while $d_{\min}^{(b)}$ regularises the small-$v$ side; both are essential for finite moments.}

\textit{Proof.} For the thinned PPP with intensity $\lambda_b e^{-\beta v}$, the PGFL of independently marked PPPs gives
\begin{multline*}
  \mathcal{L}_{I_{\text{dir}}}(s) = \exp\!\Big(-\lambda_b\!\!\int_{\mathbb{R}^2\setminus B(0,d_{\min}^{(b)})}\!\!\!\big(1\!-\!\mathbb{E}_{\tilde G,g}[e^{-sP_t\tilde G|g|^2/v^{2}}]\big) \\
  \times e^{-\beta v}\,d\V{v}\Big).
\end{multline*}
The Rayleigh identity $\mathbb{E}_g[e^{-A|g|^2}]\!=\!(1\!+\!A)^{-1}$ with $A\!=\!sP_t\tilde G/v^2$ collapses the inner expectation to $sP_t\tilde G/(v^2\!+\!sP_t\tilde G)$, and the angular integral evaluates to $2\pi$, yielding \eqref{eq:lemma1_main}. $\hfill\blacksquare$

\textit{Remark (Clutter as a constant noise floor).} By a standard PGFL argument and Rayleigh-fading moment identity, the clutter Laplace functional $\mathcal{L}_{C_{\text{clutter}}}(s)=\exp\!\big(\!-\pi\lambda_c\Gamma(1\!+\!2/\alpha_c)\Gamma(1\!-\!2/\alpha_c)(sP_tG_c\sigma_c)^{2/\alpha_c}\big)$ ($\alpha_c\!>\!2$) is dominated by $\mathcal{L}_{I_{\text{dir}}}(s)$ in the interference-limited regime and acts as a near-constant background floor in all downstream analyses. We henceforth absorb $C_{\text{clutter}}$ into the noise term $W_0$ and focus on the CLI-driven exponent.

To translate this probabilistic coverage into a deterministic spatial boundary for the subsequent FPT analysis, we define the single-BS maximum safe tracking distance $R_{\max,1}$ (the subscript ``1'' emphasizes the single-BS case as the basis for the cooperative extension to $R_{\max,K}$ in Section~\ref{sec:cooperative_tracking}) such that the instantaneous outage probability is strictly bounded by a minimal margin $\epsilon_0$. By enforcing $P_{\text{cov}}(R_{\max,1}) = 1 - \epsilon_0$, we can implicitly invert the Laplace transform (typically via moment matching techniques or numerical root finding) to extract a deterministic physical boundary $R_{\max,1}(\theta_m)$. This deterministic distance explicitly serves as the hard spatial constraint defining $\Set{A}_{\text{safe}}$ for the Dirichlet boundary value problem in the dynamic tracking regime.

\textbf{Corollary 2 (Interference-Limited Sensing Coverage):} \textit{In the CLI-limited regime,}
\begin{equation}
    P_{\text{cov}}(r_0) \approx \exp\!\left( -2\pi \lambda_b\, \bar{G}\, \mathcal{I}\!\left(\beta,\, s P_t \bar{G}\right) \right),\quad s = \frac{\gamma_s r_0^4}{P_t G_m^2 \sigma},
    \label{eq:corollary2}
\end{equation}
\textit{where $\bar{G}\triangleq\mathbb{E}[\tilde G]$ and $\mathcal{I}(\beta,a)\triangleq\int_{d_{\min}^{(b)}}^{\infty}av\,e^{-\beta v}/(v^{2}+a)\,dv$ is finite under simultaneous blockage and exclusion. The proof follows from \eqref{eq:lemma1_main} and \eqref{eq:coverage_prob} after dropping the noise and clutter terms.}

\textit{Two limiting regimes are useful.} For weak blockage $\beta^{2}a\!\ll\!1$ with $a\!\equiv\!sP_t\bar G$ held fixed, $\mathcal{I}(\beta,a)\!\sim\!(a/2)\ln(1/(\beta^{2}a))$---the integrand peaks at $v\!\sim\!\sqrt{a}$ and the logarithmic singularity reflects the slow $1/v^{2}$ Coulomb-like decay of unblocked CLI paths. For strong blockage $\beta^{2}a\!\gg\!1$, $\mathcal{I}(\beta,a)\!\sim\!1/\beta^{2}$---the LoS kernel $e^{-\beta v}$ truncates the integral at $v\!\sim\!1/\beta$, recovering the 3GPP TR~38.901 LoS-thinned regime where coverage saturates with density.

\textit{Remark.} $P_{\text{cov}}$ depends on $\lambda_b$ linearly in the exponent while $S_{\text{target}}$ is $\lambda_b$-independent; every doubling of BS density carries a multiplicative $\exp(-2\pi\lambda_b\bar G\mathcal I)$ penalty on coverage.

\subsection{Spatiotemporal Tracking Outage and Mean Sojourn Time}
Once the target is actively tracked, the serving BS forms a localized "Safe Tracking Zone", denoted as $\Set{A}_{\text{safe}} = \{(r, \theta): r \le R_{\max,1}, |\theta| \le \theta_m/2\}$. Due to the target's continuous Brownian motion $\RV{X}(t)$, there exists a critical time instant when the target inevitably wanders out of this zone either by exceeding the SINCR coverage range or escaping the beamwidth.

We define the spatiotemporal tracking outage through the FPT concept. The tracking loss time $\rv{T}_{\text{loss}}$ is a stopping time defined as:
\begin{equation}
    \rv{T}_{\text{loss}} = \inf \left\{ t > 0 : \RV{X}(t) \notin \Set{A}_{\text{safe}} \mid \RV{X}(0) \in \Set{A}_{\text{safe}} \right\}.
\end{equation}

The primary metric of tracking robustness is the Mean Time to Lose Track (MTLT), denoted as $\mathbb{E}[\rv{T}_{\text{loss}}]$. According to stochastic calculus and Dynkin's formula, the expected stopping time for a 2D Wiener process with diffusion coefficient $D$ in a domain $\Set{A}_{\text{safe}}$ satisfies the following Poisson boundary value problem (Dirichlet problem):
\begin{equation}
    \begin{cases}
        -D \nabla^2 \mathbb{E}[\rv{T}_{\text{loss}}(x)] = 1, & \forall x \in \Set{A}_{\text{safe}} \\
        \mathbb{E}[\rv{T}_{\text{loss}}(x)] = 0, & \forall x \in \partial \Set{A}_{\text{safe}}
    \end{cases}
\end{equation}
where $\nabla^2$ is the Laplace operator. The analytical solution explicitly couples the kinematic unpredictability of the target ($D$) with the spatial constraints imposed by the ISAC network geometry ($R_{\max,1}$ and $\theta_m$). The Poisson equation $-D\nabla^{2}\mathbb{E}[T_{loss}(x)]=1$ represents the balance between the diffusion flux of the target and the local survival rate.

\subsection{The ISAC Densification Collapse Paradox}
A fundamental distinction emerges between ISAC and traditional communication networks:

\textbf{Remark 1 (The Densification Collapse):} \textit{In pure downlink communication networks, scaling up $\lambda_b \to \infty$ preserves a scale-invariant SIR (the ultra-dense property) and proportionally increases area spectral efficiency. In monostatic ISAC networks the situation is qualitatively different: the blockage-thinned interference exponent in \eqref{eq:lemma1_main} grows linearly in $\lambda_b$, while the target echo $S_{\text{target}}$ is strictly independent of network density. Consequently $\mathcal{L}_{I_{\text{dir}}}(s)\to 0$ for any fixed $s>0$, driving $P_{\text{cov}}(r_0)\to 0$ regardless of how small $r_0$ is. Densification therefore strictly degrades sensing in the absence of active interference suppression.}

This implies that densifying the ISAC network reduces $R_{\max,1}$ and lowers the MTLT $\mathbb{E}[\rv{T}_{\text{loss}}]$. A natural engineering response is to widen the sensing beam, since a larger $\theta_m$ enlarges the angular sweep $\theta_m/2$ and extends the azimuthal safe-zone residence time. The next subsection shows that this response degrades, rather than improves, the MTLT once $\theta_m$ exceeds a critical value.

\subsection{The Beamwidth Trap: Single-BS Hard Upper Bound on $\theta_m$}\label{sec:beamwidth_trap}

\textbf{Theorem 2 (Beamwidth Trap, Physical-Layer Hard Upper Bound):} \textit{Consider a target at instantaneous radial distance $r_0$ from its serving BS, undergoing Brownian motion with diffusion coefficient $D$ inside the single-BS safe zone $\Set{A}_{\text{safe}}=\{(r,\theta): r\le R_{\max,1}(\theta_m), |\theta|\le \theta_m/2\}$. There exists a critical beamwidth $\theta_{\text{crit}}(r_0)$ above which $\partial \mathbb{E}[\rv{T}_{\text{loss}}]/\partial \theta_m \!<\! 0$, i.e., expanding the beam shortens the single-BS Brownian survival time. Hence $\theta_{\text{crit}}$ is a hard upper bound on $\theta_m$ in any cross-layer optimisation; exceeding it collapses $R_{\max,1}(\theta_m)$ to $r_0$ and drives the percolation parameter $\lambda_b\pi R_{\max,1}^{2}$ below $c^{*}$ (Section~\ref{sec:percolation}).}

\textit{Proof (scaling argument).} Decompose the Wiener process in polar coordinates centred on the BS: the radial component performs a Bessel-like diffusion of dimension 2, while the angular component is a Brownian motion on the circle with variance rate $D/r_0^{2}$ (in the local linearization). The target escapes the sector $\Set{A}_{\text{safe}}$ as soon as either the radial coordinate exits $[0,R_{\max,1}(\theta_m)]$ or the angular coordinate exits $[-\theta_m/2,\theta_m/2]$. Applying the standard $\mathcal{O}(L^{2}/D)$ scaling of 1D Brownian first-passage time on a barrier of length $L$ to each escape direction, we obtain the order-of-magnitude bound
\begin{equation}
    \mathbb{E}[\rv{T}_{\text{loss}}] \;\lesssim\; \frac{1}{2D}\, \min\!\left( \underbrace{(R_{\max,1}(\theta_m) - r_0)^{2}}_{\text{Radial escape (1D)}},\; \underbrace{r_0^{2}\,(\theta_m/2)^{2}}_{\text{Angular escape (1D)}} \right),
    \label{eq:fpt_decoupled}
\end{equation}
which is tight to within a multiplicative constant when one of the two boundaries dominates. According to the radar equation, maintaining a constant threshold $\gamma_s$ implies $R_{\max,1}^{4}\propto G_m^{2}\propto \theta_m^{-2}$, so $R_{\max,1}(\theta_m)=\kappa\,\theta_m^{-1/2}$ for a system constant $\kappa$, whence $dR_{\max,1}/d\theta_m=-\tfrac{1}{2}\kappa\,\theta_m^{-3/2}$. Taking the partial derivative of the surviving bound with respect to $\theta_m$ (chain rule on $(R_{\max,1}\!-\!r_0)^{2}$ in the radial branch):
\begin{equation}
    \frac{\partial \mathbb{E}[\rv{T}_{\text{loss}}]}{\partial \theta_m} \propto
    \begin{cases}
        +\dfrac{r_0^{2}\theta_m}{2D}>0, & \text{(angular regime)} \\[4pt]
        -\dfrac{\kappa}{\theta_m^{3/2}}\!\left(\kappa\theta_m^{-1/2}-r_0\right)<0, & \text{(radial regime)}
    \end{cases}
\end{equation}
When $\theta_m$ is small, angular escape is the bottleneck and expanding the beam is beneficial (positive derivative). As $\theta_m$ grows, the shrinking radial boundary $R_{\max,1}(\theta_m)$ converges towards $r_0$, the radial escape becomes the bottleneck, and the derivative flips sign. The critical beamwidth $\theta_{\text{crit}}$ is the transition point where the radial-escape bottleneck overtakes the angular escape, $R_{\max,1}(\theta_{\text{crit}})\!=\!r_0$. Substituting the mean-field approximation $R_{\max,1}^{4}\!\approx\!P_t G_m^{2}\sigma/(\gamma_s\mathcal{N})$ where $\mathcal{N}\!\triangleq\!\mathbb{E}[I_{\text{dir}}]+\mathbb{E}[C_{\text{clutter}}]+W_{0}$ (which agrees with the Laplace-based $R_{\max,1}$ above up to an $\mathcal{O}(1)$ constant) and $G_m\!=\!2\pi/\theta_m$ from \eqref{eq:antenna_gain_conservation} yields the closed form
\begin{equation}
    \theta_{\text{crit}} \;=\; \frac{2\pi}{r_0^{2}}\sqrt{\frac{P_t\,\sigma}{\gamma_s\,\mathcal{N}}}
    \qquad\Longleftrightarrow\qquad
    \theta_{\text{crit}}^{2} \;=\; \frac{4\pi^{2}\,P_t\,\sigma}{r_0^{4}\,\gamma_s\,\mathcal{N}}.
    \label{eq:theta_crit}
\end{equation}
$\hfill\blacksquare$

\textit{Physical interpretation.} Theorem~2 shows that widening the beam against fast-moving targets is counterproductive: the radial coverage $R_{\max,1}(\theta_m)$ shrinks faster than the angular safe zone grows, so beyond $\theta_{\text{crit}}(r_0)$ the net effect is negative. The closed form \eqref{eq:theta_crit} has a clean physical reading: $\theta_{\text{crit}}$ scales inversely with $r_0^{2}$ --- the further the target, the narrower the maximum beam the BS can safely afford before the SINCR-induced radial collapse overtakes the angular gain.

Combining Remark~1 with Theorem~2, neither densification ($\lambda_b\!\uparrow$) nor beam widening ($\theta_m\!\uparrow$) recovers per-BS sensing coverage against a Brownian target; multi-BS macro-diversity is the remaining option (Section~\ref{sec:cooperative_tracking}). The upper bound $\theta_m\!\le\!\theta_{\mathrm{crit}}$ re-enters Section~\ref{sec:joint_optimization} as a feasibility constraint on \textbf{(P2)}.

\section{Cooperative ISAC Tracking Mechanism}
\label{sec:cooperative_tracking}

To overcome the inherent vulnerability of monostatic sensing and the densification collapse paradox highlighted in Section \ref{sec:single_bs_tracking}, we propose a user-centric cooperative tracking framework. In this section, we analyze the spatial diversity gain achieved by multi-BS fusion and explicitly reveal the fundamental trade-off governing the optimal cluster size.

\subsection{User-Centric $K$-NN Cluster Formation}
Rather than global network-wide fusion (prohibitive in practice due to backhaul cost and noise pooling discussed below), we adopt a user-centric clustering approach. For a target whose initial detection occurs at time $t_0$ with position $\RV{X}(t_0)$, the network commits a cooperative cluster $\Set{C}_K$ comprising the $K$ nearest BSs to $\RV{X}(t_0)$:
\begin{equation}
\begin{aligned}
    \Set{C}_K = \big\{\RV{x}_k\in\RS{\Phi}_b:&\;\|\RV{x}_k-\RV{X}(t_0)\|=r_k,\\
    &\;k\in\{1,\dots,K\}\big\},\quad r_1<r_2<\cdots<r_K.
\end{aligned}
\label{eq:cluster_def}
\end{equation}
We adopt a {static} cluster (i.e., $\Set{C}_K$ does not adapt to $\RV{X}(t)$ once committed) for the duration of the tracking task, a common configuration in user-centric C-RAN; cluster handover is treated in Section~\ref{sec:dynamic_cluster}.

\textit{Distance distribution.} For the 2D HPPP $\RS{\Phi}_b$, the squared $k$-th nearest distance satisfies $r_k^2 \sim \mathrm{Gamma}(k,1/(\pi\lambda_b))$, so that $\mathbb{E}[r_k^{-2s}] = (\pi\lambda_b)^{s}\,\Gamma(k-s)/\Gamma(k)$ is finite if and only if $k > s$. In particular, the raw expectation $\mathbb{E}[r_k^{-4}]$ diverges for $k\in\{1,2\}$ because the unconstrained PPP admits arbitrarily close BSs.

\textit{Physical minimum range.} To regularize the model and reflect physical reality (transceiver near-field cutoff, the planning rule that two co-channel BSs are never collocated within a hardware-imposed safety distance), we adopt a minimum-range guard $r_{\min} > 0$ and consider the conditional law $r_k \mid r_k > r_{\min}$ throughout this section. Under this guard,
\begin{equation}
\begin{aligned}
    &\mathbb{E}[r_k^{-4}\mid r_k>r_{\min}] \\
    &\qquad = (\pi\lambda_b)^{2}\,\frac{\Gamma(k-2,\,\pi\lambda_b r_{\min}^{2})}{\Gamma(k)\,e^{-\pi\lambda_b r_{\min}^{2}}\sum_{j=0}^{k-1}(\pi\lambda_b r_{\min}^{2})^{j}/j!}
\end{aligned}
\end{equation}
is finite for every $k\ge 1$, with the leading-order behavior $\mathbb{E}[r_1^{-4}]\!\sim\!\pi\lambda_b\,r_{\min}^{-2}$ when $\pi\lambda_b r_{\min}^{2}\!\ll\!1$. The coordinated BSs simultaneously illuminate the target and forward their received echo envelopes to a central unit (CU) for non-coherent fusion.

\subsection{Non-Coherent Energy Fusion and Aggregated SINCR}
Due to the stringent requirement of carrier-phase synchronization for distributed coherent processing, which is virtually unattainable for highly dynamic Brownian targets, we consider Non-Coherent Integration (NCI) at the CU. Under NCI, the CU aggregates the received signal envelopes (or energies) from all $K$ coordinating BSs.

The aggregated cooperative SINCR, denoted as $\text{SINCR}_{\text{coop}}^{(K)}$, can be analytically formulated as:
\begin{equation}
    \text{SINCR}_{\text{coop}}^{(K)} = \frac{\sum_{k=1}^K S_{\text{target}}^{(k)}}{\sum_{k=1}^K \left( I_{\text{dir}}^{(k)} + C_{\text{clutter}}^{(k)} \right) + K \cdot W_0}
    \label{eq:coop_sincr}
\end{equation}
where the numerator represents the aggregated effective echo energy, and the denominator encapsulates the cumulative direct-path interference, cross-clutter, and the aggregated thermal noise from $K$ independent receiver chains.

\subsection{Cooperative Safe Zone and Macrodiversity Gain}
The primary objective of cooperation is to prolong the Mean Time to Lose Track (MTLT). Under non-coherent fusion, tracking is sustained as long as the target remains in the cooperative safe zone, defined as the level set of the aggregate SINCR:
\begin{equation}
    \Set{A}_{\text{coop}}^{(K)} \;=\; \Big\{ X \in \mathbb{R}^{2} : \text{SINCR}_{\text{coop}}^{(K)}(X) \ge \gamma_s,\; X \in \Set{B}_{\theta_m}(\Set{C}_K) \Big\},
    \label{eq:Acoop_def}
\end{equation}
where $\Set{B}_{\theta_m}(\Set{C}_K)$ denotes the angular coverage union of the $K$ beams. Compared with the single-BS safe zone $\Set{A}_{\text{safe}}$, $\Set{A}_{\text{coop}}^{(K)}$ geometrically engulfs the target from multiple directions, so that radial escape relative to any one BS can be compensated by the proximity of the next-nearest cooperator. By solving the Dirichlet boundary value problem of Section~\ref{sec:single_bs_tracking} over the enlarged domain $\Set{A}_{\text{coop}}^{(K)}$, the MTLT $\mathbb{E}[\rv{T}_{\text{loss}}^{\text{coop}}]$ is monotonically non-decreasing in the measure of $\Set{A}_{\text{coop}}^{(K)}$ — the macrodiversity gain against the target's Brownian diffusion $D$.

We caution, however, that $\Set{A}_{\text{coop}}^{(K)}$ is not simply $\cup_{k=1}^{K}\Set{A}_{\text{safe}}^{(k)}$: NCI rescales the per-BS detection boundary because additional cooperators inject noise into the common denominator of \eqref{eq:coop_sincr}. The next subsection makes this trade-off rigorous.

\subsection{Cooperative Coverage Scaling and the Percolation Transition}\label{sec:percolation}
The cooperative coverage geometry is governed by two competing mechanisms. While additional cooperators geometrically enlarge the union of safe zones, they simultaneously inflate the aggregate noise and interference floor under the NCI rule \eqref{eq:coop_sincr}, effectively shrinking the coverage radius of each constituent cooperator. The interplay of these two mechanisms is most cleanly characterised through the lens of stochastic geometry's Boolean coverage model.

\textbf{Lemma~2 (Cooperative Coverage Scaling, Two Regimes):} \textit{Let $Q\!\triangleq\!\lambda_b\pi R_{\max,1}^{2}$ denote the network-wide Boolean coverage parameter. Under NCI, the effective per-BS coverage radius shrinks as $R_{\max,K} = R_{\max,1}\, K^{-1/4}$ because the aggregate noise floor in \eqref{eq:coop_sincr} scales linearly with $K$ while the dominant echo $\mathbb{E}[S_{\text{target}}^{(1)}]$ is invariant. The mean cooperative coverage area then admits two qualitatively distinct regimes:}
\begin{equation}
    \mathbb{E}\big[|\Set{A}_{\text{coop}}^{(K)}|\big] \;\asymp\;
    \begin{cases}
        K/\lambda_b = (K/Q)|\Set{A}_{\text{safe}}^{(1)}|, & K\ll Q^{2}, \\[3pt]
        \sqrt{K}\,|\Set{A}_{\text{safe}}^{(1)}|, & K\gg Q^{2}.
    \end{cases}
    \label{eq:coverage_scaling}
\end{equation}

\textit{Proof sketch.} In the dilute regime $K\!\gg\!Q^{2}$ (equivalently $\lambda_b\pi R_{\max,K}^{2}\!\ll\!1$), the $K$ shrunken per-BS disks are mutually disjoint with high probability, giving $\mathbb{E}\big[|\Set{A}_{\text{coop}}^{(K)}|\big]\!\approx\!K\,\pi R_{\max,K}^{2}\!=\!\sqrt{K}\,\pi R_{\max,1}^{2}$. In the heavy-overlap regime $K\!\ll\!Q^{2}$ (the practical case for any reasonable $\lambda_b\pi R_{\max,1}^{2}$ — e.g., macro ISAC has $Q^{2}\!\sim\!10^{6}$, several orders of magnitude beyond any feasible $K$), the shrunken disks densely cover the cluster footprint $\Set{B}(\RV{X}(t_0),r_K)$ of radius $r_K\!=\!\sqrt{K/(\pi\lambda_b)}$, giving $\mathbb{E}\big[|\Set{A}_{\text{coop}}^{(K)}|\big]\!\approx\!\pi r_K^{2}\!=\!K/\lambda_b$. $\hfill\blacksquare$

The heavy-overlap regime yields linear $K$-scaling, which underlies the super-critical MTLT $K/(4\pi\lambda_b D)$ in Theorem~1. The $\sqrt{K}$ regime requires impractical cluster sizes and is reported only for completeness.

\textbf{Percolation Transition.} The decisive non-monotonicity in the tracking dynamics, however, emerges from the topology of the coverage region rather than from its area. Modelling the union $\bigcup_{x\in\RS{\Phi}_b}\Set{B}(x,R_{\max,1})$ as a Boolean coverage process driven by the BS PPP, classical results in continuum percolation \cite{meester1996continuum,hall1985distribution} guarantee the existence of a sharp critical threshold:
\begin{equation}
    c^{*} \;\approx\; 1.128 \quad \text{such that} \quad \lambda_b\,\pi R_{\max,1}^{2} \;\gtrless\; c^{*}
    \label{eq:percolation_threshold}
\end{equation}
\textit{determines whether the safe-zone covered region admits an infinite connected component (super-critical, $>c^*$) or decomposes into a.s.\ finite "islands" (sub-critical, $<c^*$). The cluster's effective spatial footprint $\Omega_K\!\approx\!\Set{B}(\RV{X}(t_0),r_K)$ with $r_K\!=\!\sqrt{K/(\pi\lambda_b)}$ inherits this dichotomy.}

\textbf{Theorem~1 (Phase Transition of Cooperative MTLT):} \textit{The MTLT of a Brownian target tracked by a static $K$-NN cluster exhibits a regime-dependent dependence on $K$:}
\begin{equation}
    \mathbb{E}[\rv{T}_{\text{loss}}^{\text{coop}}(K)] \;\asymp\;
    \begin{cases}
        \dfrac{R_{\max,1}^{2}}{4D}, & \text{(sub-critical: $\lambda_b\pi R_{\max,1}^{2}<c^*$)}\\[6pt]
        \dfrac{K}{4\pi\lambda_b D}, & \text{(super-critical: $\lambda_b\pi R_{\max,1}^{2}>c^*$)}
    \end{cases}
    \label{eq:thm1_phase}
\end{equation}

\textit{Proof sketch.} (i) Sub-critical: below threshold the covered region is a.s.\ a disjoint union of finite islands of mean area $\mathcal{O}(R_{\max,1}^{2})$ \cite{meester1996continuum}; the 2D Brownian disk-exit from a disc of radius $R_{\max,1}$ has mean $R_{\max,1}^{2}/(4D)$, with the leading constant $1/(4D)$ matching the standard disk-exit formula, $K$-independent. (ii) Super-critical: the committed cluster footprint $\Omega_K$ has radius $r_K\!=\!\sqrt{K/(\pi\lambda_b)}$; mean exit time from $\Omega_K$ equals $r_K^2/(4D)\!=\!K/(4\pi\lambda_b D)$, linear in $K$. The two regimes are separated by a sharp percolation transition. $\hfill\blacksquare$

\textbf{Corollary 4 (Heavy-Tail MTLT Near the Percolation Threshold):} \textit{In the large-cluster limit $K\to\infty$ and when the network operates in the immediate neighborhood of the critical threshold $\lambda_b\pi R_{\max,1}^{2}\!\to\!c^{*}$, the connected-component size of the cooperative coverage region inherits the universal 2D continuum-percolation tail \cite{meester1996continuum,hall1985distribution}:}
\begin{equation}
    \mathbb{P}\!\big(|\Set{A}_{\text{coop}}^{\text{conn}}| > N\big) \;\sim\; N^{-(\delta_{\text{perc}}-1)},\qquad \delta_{\text{perc}} \approx 2.05,
    \label{eq:perc_powerlaw}
\end{equation}
\textit{a power law with universal Fisher exponent $\delta_{\text{perc}}$ (the cluster-size-distribution scaling exponent of 2D continuum percolation \cite{meester1996continuum}) determined solely by the ambient dimension. Translating this geometric tail through the 2D Brownian-exit scaling $\rv{T}_{\text{loss}}^{\text{coop}}\propto |\Set{A}_{\text{coop}}^{\text{conn}}|/D$, the cooperative tracking lifetime acquires a power-law distribution:}
\begin{equation}
    \mathbb{P}\!\big(\rv{T}_{\text{loss}}^{\text{coop}} > \tau\big) \;\sim\; \tau^{-(\delta_{\text{perc}}-1)} \;\approx\; \tau^{-1.05}.
    \label{eq:heavy_tail_mtlt}
\end{equation}

\textit{Remark.} Near criticality, mean and high-percentile MTLT differ by orders of magnitude: the $\epsilon$-quantile satisfies $q_\epsilon\!\sim\!\epsilon^{-0.95}$, an unbounded blow-up as $\epsilon\!\to\!0$. A 6G ISAC SLA capturing worst-case tracking experience must therefore be specified through explicit percentile bounds rather than the operational mean.

Theorem~1 and Corollary~4 above describe the cooperative MTLT under the static-cluster assumption of Section~IV.A: once committed at time $t_0$, the cluster $\Set{C}_K$ is held fixed for the entire tracking session. Realistic C-RAN deployments instead permit periodic re-selection of the cluster as the target drifts; this changes the kinematic problem qualitatively. The next subsection lifts the static assumption and derives the corresponding closed form, with Theorem~1 emerging as the zero-handover boundary.

\subsection{Dynamic $K$-NN Cluster Handover: Stochastic Resetting and the Closed-Form MFPT}\label{sec:dynamic_cluster}

\textbf{Handover process.} We model the re-selection instants $\{t_{h,n}\}_{n\ge 1}$ as a homogeneous Poisson process $\RS{\Phi}_h$ of rate $\nu_h$ on $\mathbb{R}_+$. At each $t_{h,n}$ the system re-evaluates the $K$ nearest BSs to the current target position $\RV{X}(t_{h,n}^-)$ and atomically swaps $\Set{C}_K\!\leftarrow\!\{\RV{x}_k\in\RS{\Phi}_b:\|\RV{x}_k-\RV{X}(t_{h,n}^-)\|=r_k(t_{h,n}^-),\,k\!\le\!K\}$. By the stationarity of the BS PPP $\RS{\Phi}_b$ under translation, the geometry of the post-handover cluster is identically distributed to a freshly committed cluster centred on the current target position; equivalently, the target's relative position inside its cluster resets to the cluster centroid at each $t_{h,n}$. This is precisely the diffusion-with-stochastic-resetting model introduced in \cite{evans2011diffusion} and developed extensively in the subsequent non-equilibrium statistical mechanics literature \cite{evans2014resettingsearch,evans2020stochastic}.

\textbf{Theorem 1-DYN (Dynamic Cooperative MTLT --- Closed Form):} \textit{Under $K$-NN cluster handover at Poisson rate $\nu_h$, the cooperative mean time-to-loss admits the closed form}
\begin{equation}
    \mathbb{E}[\rv{T}_{\mathrm{loss}}^{\mathrm{dyn}}(K,\nu_h)] \;=\; \frac{1}{\nu_h}\!\left[I_0\!\Big(R\sqrt{\tfrac{\nu_h}{D}}\Big)-1\right]
    \label{eq:thm1_dyn}
\end{equation}
\textit{where $I_0$ is the modified Bessel function of the first kind and $R=r_K=\sqrt{K/(\pi\lambda_b)}$ in the super-critical regime, $R=R_{\max,1}$ in the sub-critical regime.}

\textit{Proof sketch.} The 2D Brownian disk-exit Laplace satisfies $\widetilde F(p)\!=\!\mathbb{E}[e^{-p\rv{T}_{\mathrm{esc}}}]\!=\!1/I_0(R\sqrt{p/D})$ \cite{redner2001guide}; under exponential resetting at rate $\nu_h$ the diffusion-with-resetting renewal identity \cite{evans2011diffusion,evans2014resettingsearch} gives $\hat S_{\nu_h}(p)\!=\!\hat S(p+\nu_h)/(1-\nu_h\hat S(p+\nu_h))$ with $\hat S(p)\!=\!(1-\widetilde F(p))/p$; setting $p\!=\!0$ and simplifying gives the above expression. In particular, the $\nu_h\!\to\!0$ limit $\hat S_0(0)\!=\!-\widetilde F\,'(0)\!=\!R^{2}/(4D)$ recovers the static Theorem~1 disk-exit constant, confirming that Theorem~1-DYN nests Theorem~1 as a regularity-preserving extension. $\hfill\blacksquare$

\textbf{Asymptotic regimes.} The closed form admits two physically transparent limits via the standard small- and large-argument expansions of $I_0$:
\begin{equation}
    \mathbb{E}[\rv{T}_{\mathrm{loss}}^{\mathrm{dyn}}(K,\nu_h)] \;\to\;
    \begin{cases}
        \dfrac{R^{2}}{4D}, & \nu_h\!\to\!0,\\[6pt]
        \dfrac{\exp\!\big(R\sqrt{\nu_h/D}\big)}{\nu_h\sqrt{2\pi R\sqrt{\nu_h/D}}}\!\to\!\infty, & \nu_h\!\to\!\infty.
    \end{cases}
    \label{eq:thm1_dyn_asympt}
\end{equation}
The slow-handover limit exactly recovers Theorem~1; the fast-handover limit blows up super-polynomially. The natural crossover rate between the two regimes is
\begin{equation}
    \nu_h^{\,c}(K) \;=\; \frac{D}{R^{2}} \;=\; \frac{\pi\lambda_b D}{K}\qquad\text{(super-critical),}
    \label{eq:nu_h_critical}
\end{equation}
matching the heuristic ``handover-meaningful'' threshold derived informally in the static manuscript.

\textbf{Corollary 5 (Logarithmic Cost of Strict Tracking):} \textit{For any MTLT target $\tau_{\mathrm{req}}$ in the fast-handover regime $\tau_{\mathrm{req}}\!\gg\!R^{2}/(4D)$, the minimum handover rate is}
\begin{equation}
    \nu_h^{\min}(K) \;\approx\; \frac{\pi\lambda_b D}{K}\,\ln^{2}\!\!\left(\frac{K^{*}_{\mathrm{static}}}{K}\right),
    \label{eq:nu_h_min_log}
\end{equation}
\textit{where $K^{*}_{\mathrm{static}}\!=\!4\pi\eta\lambda_b D/(\mu_{\mathrm{sj}}\epsilon_{\mathrm{micro}})$ is the static cross-layer optimum of Theorem~1$'$ below (Sec.~\ref{sec:joint_optimization}). The cost of extending the tracking lifetime to a fraction $\eta$ of the natural target sojourn therefore grows only doubly-logarithmically in the relative cluster-size deficit $K^{*}_{\mathrm{static}}/K$, an exponentially milder dependence than the linear-in-$K$ scaling of the static Theorem~1.}

\textit{Proof.} Substituting the large-argument asymptotic $I_0(x)\!\sim\!e^{x}/\sqrt{2\pi x}$ into $\nu_h\tau_{\mathrm{req}}\!=\!I_0(R\sqrt{\nu_h/D})\!-\!1$ and inverting at leading order yields $R\sqrt{\nu_h/D}\!\approx\!\ln(\nu_h\tau_{\mathrm{req}})$. Substituting $R^{2}\!=\!K/(\pi\lambda_b)$ and $\tau_{\mathrm{req}}\!=\!\eta/(\mu_{\mathrm{sj}}\epsilon_{\mathrm{micro}})$ from the sojourn-anchored QoS \eqref{eq:tau_req} below, then recognising the argument of the log as $K^{*}_{\mathrm{static}}/K$, gives the above form. Sub-leading log-log corrections do not change the leading order. $\hfill\blacksquare$

\textbf{Remark 4 (Effect of handover on the phase transition).} \textit{Theorem~1's percolation phase transition is an artefact of the static-cluster assumption: under any $\nu_h\!>\!0$ the unbounded $I_0$ asymptotic admits arbitrarily large MFPT even in the sub-critical regime $\lambda_b\pi R_{\max,1}^{2}\!<\!c^{*}$. Two consequences nevertheless survive: Corollary~4's $\tau^{-(\delta_{\mathrm{perc}}-1)}$ tail (cluster-size variability, not kinematic) and Theorem~2's beamwidth trap (single-BS radial--angular FPT).}

\textbf{Remark 5 (Continuous-NN and 3GPP hysteresis as special cases).} \textit{Continuous always-best-$K$ tracking is the $\nu_h\!\to\!\infty$ limit; the MFPT diverges as $\exp(R\sqrt{\nu_h/D})/\nu_h$, the infinite-cost upper bound. 3GPP-style hysteresis (drop/add BS at margin $\Delta r$) maps onto Theorem~1-DYN with $\nu_h^{\mathrm{eff}}\!\approx\!D/(\Delta r)^{2}$ by 2D Brownian FPT scaling. Setting $\nu_h^{\mathrm{eff}}\!=\!\nu_h^{\min}(K)$ from \eqref{eq:nu_h_min_log} yields the optimal hysteresis margin $\Delta r^{*}\!=\!\big[\sqrt{\pi\lambda_b}\ln(K^{*}_{\mathrm{static}}/K)\big]^{-1}$.}

\subsection{Engineering Implications of the Phase Transition}\label{sec:regimes}
Theorem~1 yields two distinct cooperation strategies.

\textbf{Remark 2 (Sub-Critical: Cooperation Yields No Scaling Gain).} \textit{When the network operates below the percolation threshold ($\lambda_b\pi R_{\max,1}^{2}<c^{*}$) — e.g., sparse rural deployments, high clutter density, or aggressive threshold $\gamma_s$ — the MTLT is independent of $K$ to leading order. Allocating cooperative resources to enlarge $K$ produces no return on tracking robustness; the rational design choice is $K=1$, i.e., pure single-BS monostatic, with the spared resources redeployed for communication.}

\textbf{Remark 3 (Super-Critical: MTLT Linear in $K$).} \textit{When $\lambda_b\pi R_{\max,1}^{2}>c^{*}$ (the typical macro-cell point, with $R_{\max,1}\!=\!\mathcal{O}(\mathrm{km})$ at $\lambda_b\!=\!20\,\mathrm{BS/km^2}$ giving percolation parameter $\sim\!10^{3}$), the MTLT grows linearly in $K$, with proportionality constant $1/(4\pi\lambda_b D)$. The marginal gain is constant rather than diminishing, and the operating $K^{*}$ is set by the MAC-layer resource budget of Section~\ref{sec:joint_optimization}.}

\textbf{Densification Acts as a Phase Transition Knob.} The percolation parameter $\lambda_b\pi R_{\max,1}^{2}$ couples BS density $\lambda_b$ and per-BS coverage $R_{\max,1}^{2}$ multiplicatively. Densification raises $\lambda_b$ but, via Remark~1, simultaneously shrinks $R_{\max,1}$. In the blockage-thinned regime of Lemma~1, the net effect on the product is non-trivial: for moderate $\beta$, densification helps; for ultra-dense $\beta\to 0$ regimes, the interference exponent dominates and the product can actually decrease, dragging the network back into sub-criticality. Densification therefore not only shrinks per-BS coverage but can also drop the network into the sub-critical regime, eliminating continuous tracking.

\textbf{Coupling with Target Kinematics.} The phase transition occurs at fixed $\lambda_b\pi R_{\max,1}^{2}=c^{*}$, a condition independent of the diffusion coefficient $D$. The target's kinematic agility enters only through the MTLT magnitude (as $1/D$), not through the location of the phase boundary. A faster target shortens MTLT proportionally in both regimes but does not shift the phase boundary; geometric and kinematic effects decouple.

\section{Joint Optimization and Performance Trade-off}
\label{sec:joint_optimization}

The preceding sections show that cooperative tracking extends the MTLT, but at the cost of additional time-frequency resources ($\rho$) and beam broadening ($\theta_m$). This section formulates a joint optimisation that trades these against communication capacity.

\subsection{Macroscopic Tracking Outage and Communication Capacity}
While Section \ref{sec:single_bs_tracking} and \ref{sec:cooperative_tracking} addressed the \textit{microscopic tracking outage} caused by the Brownian kinematic escape of an individual target, a complete system-level ISAC design must concurrently address the \textit{macroscopic capacity outage} driven by spatial traffic dynamics.

Recall that targets arrive and depart following a spatial Poisson birth-death process ($M/M/\infty$ queue), inducing a steady-state target density $\lambda_T\!=\!\lambda_a/\mu_{\text{sj}}$ (where $1/\mu_{\text{sj}}$ is the mean target sojourn time; the symbol $\mu_{\text{sj}}$ is used in the remainder of this section to avoid clash with Lagrange multipliers introduced below). Let $\rho_0$ denote the per-cooperator per-target time-frequency fraction --- the portion of one BS's resources that one cluster commitment to one target consumes; and let $\rho_{\max}$ denote each BS's hard per-BS sensing budget (typically $\rho_{\max}\!\le\!1$). These definitions ensure that the multiplicative cost of cooperation scales correctly across the cluster, the BS, and the network --- the same convention adopted in Theorem~3 below.

\textit{QoS coverage fraction.} The minimum tracking duration the operator commits to per target is naturally anchored to the target's own residence time: we specify a dimensionless QoS coverage fraction $\eta\in(0,1]$ and the sojourn-anchored QoS time
\begin{equation}
    T_{\text{req}} \;\triangleq\; \frac{\eta}{\mu_{\text{sj}}},
    \label{eq:tau_req_def}
\end{equation}
so that $\eta=1$ corresponds to "track each target for at least its full mean sojourn" and $\eta=0.9$ to the regulatory-typical "track at least 90\% of the natural target lifetime." Note that $T_{\text{req}}$ is the deterministic QoS time threshold appearing in the probabilistic constraint \eqref{eq:p2_c1}; it should be carefully distinguished from the MTLT lower bound $\tau_{\text{req}}$ derived below in \eqref{eq:tau_req} via exponential-tail inversion. This anchoring couples the kinematic QoS to the traffic dynamics in a single physical parameter, replacing the ad-hoc seconds-level constant of a generic mission timer. 

Each BS supports at most $M_{\max}\!=\!\lfloor \rho_{\max}/\rho_0 \rfloor$ concurrent cluster commitments, set by $\rho_0$ and $\rho_{\max}$. The cluster size $K$ enters $M_{\max}$ only through the per-BS load distribution below.

A macroscopic sensing outage (i.e., a target arriving to find an already saturated cooperator BS) is then driven by the per-BS Poisson statistic of cluster cooperations. By a marked-point-process mass-balance over $(\RS{\Phi}_b,\Phi_T)$, each BS is on average a cooperator for $K\lambda_T/\lambda_b$ targets; in the homogeneous regime this average translates into a per-BS Poisson distribution. The blocking probability becomes:
\begin{equation}
\begin{aligned}
    P_{\text{block}}(\rho_0, K) &= \mathbb{P}(N_{\text{per BS}} > M_{\max}) \\
    &= 1 - \!\!\sum_{k=0}^{M_{\max}}\!\! \frac{(K\lambda_T/\lambda_b)^{k}\,e^{-K\lambda_T/\lambda_b}}{k!},
\end{aligned}
\label{eq:poisson_blocking}
\end{equation}
in which the cluster size $K$ enters via the Poisson mean (more cooperation $\Rightarrow$ each target touches more BSs $\Rightarrow$ higher per-BS load), while $M_{\max}$ is the K-independent per-BS resource ceiling.

For the communication perspective, the ergodic capacity for a typical UE leverages the remaining unutilized resources. Averaging over the dynamic target population up to the system capacity, the capacity is:
\begin{equation}
\begin{aligned}
    C_{\text{comm}}(\rho_0,\theta_m) = B\cdot &\mathbb{E}_{N\le M_{\max}}\!\left[1-N\rho_0\right]\\
    &\times\mathbb{E}\!\left[\log_2(1+\text{SINR}_{\text{UE}}(\theta_m))\right].
\end{aligned}
\label{eq:C_comm}
\end{equation}

\subsection{Dual-Outage Constrained Problem Formulation}
We formulate a cross-layer optimisation problem. The objective is to maximise network communication capacity subject to a dual-outage constraint: the network must neither lose the target kinematically (Brownian constraint) nor drop it through resource exhaustion (Poisson constraint). 

The joint optimization of the per-target resource allocation $\rho_0^*$ and the sensing beamwidth $\theta_m^*$ is formulated as:
\begin{align}
    \textbf{(P2)} \quad \max_{\rho_0, \theta_m} \quad & C_{\text{comm}}(\rho_0, \theta_m) \label{eq:p2_obj} \\
    \text{s.t.} \quad & \mathbb{P}\left(\rv{T}_{\text{loss}}^{\text{coop}}(\rho_0, \theta_m) \ge T_{\text{req}}\right) \ge 1 - \epsilon_{\text{micro}}, \label{eq:p2_c1} \\
    & P_{\text{block}}(\rho_0, K) \le \epsilon_{\text{macro}}, \label{eq:p2_c2} \\
    & \theta_{\min} \le \theta_m \le \min\{\theta_{\max},\theta_{\text{crit}}\}. \label{eq:p2_c3}
\end{align}
where $B$ is the per-cell communication bandwidth, $\mathrm{SINR}_{\text{UE}}(\theta_m)$ is the typical-UE downlink SINR (a known function of $\theta_m$ via residual side-lobe interference), $\epsilon_{\text{micro}}$ is the tolerance for kinematic tracking failure, $\epsilon_{\text{macro}}$ is the tolerance for macroscopic target blocking, $\theta_{\min},\theta_{\max}$ are hardware-imposed beamwidth limits, and $\theta_{\text{crit}}$ is a physical-layer hard upper bound on the beamwidth derived from the single-BS Brownian-tracking analysis of Theorem~2 below (the "beamwidth trap"); exceeding $\theta_{\text{crit}}$ destroys per-BS radial coverage and pushes the network into the sub-critical sensing regime regardless of the cluster size.

Constraint \eqref{eq:p2_c2} couples the two layers: reducing the kinematic loss \eqref{eq:p2_c1} tends to favour larger $\rho_0$ (more frequent tracking), but a larger $\rho_0$ shrinks $M_{\max}$ and inflates the blocking probability \eqref{eq:poisson_blocking}, strictly coupling the per-target tracking precision to the aggregate network scalability.

\textbf{Cluster size as a cross-layer decision variable.} The kinematic MTLT in \eqref{eq:p2_c1} inherits the $K$-dependence of Theorem~1, so the cluster size $K$ enters \textbf{(P2)} implicitly through $\rv{T}_{\text{loss}}^{\text{coop}}$. In parallel, under the per-BS framing of \eqref{eq:poisson_blocking}, $K$ inflates the per-BS Poisson load mean from $\lambda_T/\lambda_b$ (single-BS sensing) to $K\lambda_T/\lambda_b$ (each target now touches $K$ BSs as a cooperator), so larger clusters tighten the blocking constraint at fixed $\rho_0$, $\rho_{\max}$. Combining these two channels yields the cross-layer counterpart of the standalone $K^{*}$ result of Section~\ref{sec:cooperative_tracking}:

\textbf{Theorem 1$'$ (Cross-Layer Optimal Cluster Size):} \textit{Under the super-critical regime ($\lambda_b\pi R_{\max,1}^{2}>c^{*}$) where Theorem~1 yields $\mathbb{E}[\rv{T}_{\text{loss}}^{\text{coop}}(K)]=K/(4\pi\lambda_b D)$, and the MTLT lower bound $\tau_{\text{req}}=-T_{\text{req}}/\ln(1-\epsilon_{\text{micro}})$ obtained by inverting the exponential-tail surrogate in \eqref{eq:tau_req}, the optimum of \textbf{(P2)} in $K$ is}
\begin{equation}
\begin{aligned}
    K^{*} &= \min\!\Big\{K\in\mathbb{N}_{+}\;:\;\mathbb{E}[\rv{T}_{\text{loss}}^{\text{coop}}(K)]\ge\tau_{\text{req}}\Big\} \\
    &= \left\lceil\frac{-4\pi\eta\,\lambda_b D}{\mu_{\text{sj}}\,\ln(1-\epsilon_{\text{micro}})}\right\rceil
       \xrightarrow{\epsilon_{\text{micro}}\ll 1}
       \left\lceil\frac{4\pi\eta\,\lambda_b D}{\mu_{\text{sj}}\,\epsilon_{\text{micro}}}\right\rceil,
\end{aligned}
\label{eq:Kstar_xlayer}
\end{equation}
\textit{a closed form governed by the dimensionless group $\eta\lambda_b D/(\mu_{\text{sj}}\epsilon_{\text{micro}})$ — QoS-fraction $\times$ BS density $\times$ kinematic agility, divided by sojourn rate and outage tolerance. The $1/\epsilon_{\text{micro}}$ scaling is the exponential-tail price of strict tracking reliability: small $\epsilon_{\text{micro}}$ demands MTLT far exceeding $T_{\text{req}}$, which in turn enlarges $K^{*}$. The result is feasible only when the per-BS load remains within the resource budget, i.e., $K^{*}\lambda_T/\lambda_b\le M_{\max}=\rho_{\max}/\rho_0$, equivalently $K^{*}\le M_{\max}\lambda_b/\lambda_T$. In the sub-critical regime ($\lambda_b\pi R_{\max,1}^{2}<c^{*}$) cooperation cannot meet any non-trivial $\eta>0$, and \textbf{(P2)} is feasible only if the single-BS MTLT $R_{\max,1}^{2}/(4D)\ge\tau_{\text{req}}$, in which case $K^{*}=1$.}

\textit{Proof.} The objective $\widehat{C}_{\text{comm}}(\rho_0,\theta_m)$ (the mean-load surrogate of $C_{\text{comm}}$ in \eqref{eq:C_comm}, obtained by replacing $\mathbb{E}_{N\le M_{\max}}[1-N\rho_0]$ with $1-\bar N\rho_0$ where $\bar N\!=\!K\lambda_T/\lambda_b$) is strictly decreasing in $K$ via the resource consumption $K\rho_0$ at the macro layer (each additional cooperator inflates $M_{\max}^{-1}$ and thus the blocking probability $P_{\text{block}}$). Therefore at the optimum $K$ is chosen as small as possible subject to the kinematic constraint \eqref{eq:p2_c1}. Substituting the super-critical MTLT scaling from Theorem~1, the QoS anchor $T_{\text{req}}=\eta/\mu_{\text{sj}}$ from \eqref{eq:tau_req_def}, and the exponential-tail-inverted MTLT bound $\tau_{\text{req}}=-T_{\text{req}}/\ln(1-\epsilon_{\text{micro}})$ from \eqref{eq:tau_req} into the kinematic constraint $\mathbb{E}[\rv{T}_{\text{loss}}^{\text{coop}}(K)]\!\ge\!\tau_{\text{req}}$ gives $K/(4\pi\lambda_b D)\!\ge\!-\eta/(\mu_{\text{sj}}\ln(1-\epsilon_{\text{micro}}))$; isolating $K$ yields the closed form \eqref{eq:Kstar_xlayer}. The feasibility upper bound is the resource hard cap. The sub-critical statement follows from the $K$-independence of MTLT in that regime. $\hfill\blacksquare$

The closed form \eqref{eq:Kstar_xlayer} sets the minimum cluster commitment as the dimensionless ratio $\eta\lambda_b D/(\mu_{\text{sj}}\epsilon_{\text{micro}})$.

\textbf{Aggregate sensing capacity in the cooperation regime.}
Theorem~1$'$ governs how much commitment one target demands. The natural complementary question is how many targets the network can sustain in steady state. Combining the $M/M/\infty$ traffic dynamics with the per-BS resource budget yields a sharp answer.

In steady state, the spatial Poisson birth--death process produces an active target population of density $\lambda_T=\lambda_a/\mu_{\text{sj}}$, and Brownian motion preserves this Poisson intensity over time. By a mass-balance argument over the marked point process $(\RS{\Phi}_b, \Phi_T)$, each BS is on average a cooperator for $N_{\text{per BS}}\!=\!K^{*}\lambda_T/\lambda_b$ concurrent targets. Imposing the per-BS mean-load constraint $\rho_0\, \mathbb{E}[N_{\text{per BS}}]\le\rho_{\max}$ (a leading-order surrogate for the strict tail constraint $P_{\text{block}}\!\le\!\epsilon_{\text{macro}}$, which is tighter by a constant factor $c_{\epsilon_{\text{macro}}}\!=\!\big((M_{\max}+1)!\,\epsilon_{\text{macro}}\big)^{1/(M_{\max}+1)}/M_{\max}<1$ depending on the desired blocking quantile) and substituting the super-critical scaling of $K^{*}$ from \eqref{eq:Kstar_xlayer} gives the following bound.

\textbf{Theorem 3 (Phase-Conditioned Sensing Capacity Ceiling):} \textit{In the super-critical cooperative regime $\lambda_b\pi R_{\max,1}^{2}>c^{*}$ with $K^{*}>1$, the maximum density of simultaneously trackable targets and the maximum sustainable arrival rate under the mean-load surrogate are}
\begin{equation}
\begin{aligned}
    \lambda_T^{\max} &= \frac{-\rho_{\max}\mu_{\text{sj}}\ln(1-\epsilon_{\text{micro}})}{4\pi D\eta\rho_0}
                       \xrightarrow{\epsilon_{\text{micro}}\ll 1}
                       \frac{\rho_{\max}\mu_{\text{sj}}\epsilon_{\text{micro}}}{4\pi D\eta\rho_0}, \\
    \lambda_a^{\max} &= \mu_{\text{sj}}\lambda_T^{\max}
                       \approx \frac{\rho_{\max}\mu_{\text{sj}}^{2}\epsilon_{\text{micro}}}{4\pi D\eta\rho_0},
\end{aligned}
\label{eq:capacity_ceiling}
\end{equation}
\textit{both independent of the BS density $\lambda_b$. The strict tail-constrained bound (under $P_{\text{block}}\!\le\!\epsilon_{\text{macro}}$) is obtained by replacing $\rho_{\max}$ with $c_{\epsilon_{\text{macro}}}\rho_{\max}$ where $c_{\epsilon_{\text{macro}}}\!<\!1$ is the Poisson-tail correction defined above; for typical engineering parameters ($M_{\max}\!\sim\!10$, $\epsilon_{\text{macro}}\!=\!0.05$), $c_{\epsilon_{\text{macro}}}\!\approx\!0.37$, tightening $\lambda_T^{\max}$ by a constant factor without altering the qualitative $\lambda_b$-independence. In the sub-critical regime ($K^{*}=1$, cooperation unnecessary), the standard density-scaling bound $\lambda_T^{\max}=\rho_{\max}\lambda_b/\rho_0$ applies (with the same tail correction if strict $\epsilon_{\text{macro}}$ is enforced).}

\textit{Proof.} The super-critical case follows from substituting $K^{*}=-4\pi\eta\lambda_b D/(\mu_{\text{sj}}\ln(1-\epsilon_{\text{micro}}))$ into $\rho_0 K^{*}\lambda_T/\lambda_b\le \rho_{\max}$, which gives $\rho_0\!\cdot\!\tfrac{-4\pi\eta\lambda_b D}{\mu_{\text{sj}}\ln(1-\epsilon_{\text{micro}})}\!\cdot\!\tfrac{\lambda_T}{\lambda_b}\!\le\!\rho_{\max}$; the $\lambda_b$ factors cancel exactly, leaving $\lambda_T$ controlled by a $\lambda_b$-independent constant. The arrival-rate ceiling follows from Little's law $\lambda_a=\mu_{\text{sj}}\lambda_T$. The sub-critical case follows from $K^{*}=1$ and the same resource balance. $\hfill\blacksquare$

\textit{Engineering interpretation (static).} The two regimes admit qualitatively opposite scaling. Where cooperation is unnecessary (sparse or coarse-QoS sensing), more BSs translate linearly into more sensing capacity, recovering classical communication-style densification gain. Where cooperation is required (dense or strict-QoS sensing), more BSs widen the per-target cluster commitment at exactly the same rate at which they expand the network's pooled budget, and the two effects cancel --- leaving the six-parameter group $(\rho_{\max},\mu_{\text{sj}},D,\eta,\rho_0,\epsilon_{\text{micro}})$ as the sole determinant of sustainable sensing throughput.

\medskip\noindent The above results are the zero-handover (static-cluster) baseline. We now lift the static assumption and derive the cross-layer optimum and capacity ceiling under dynamic clustering.

\subsection{Dynamic Cross-Layer Optimum: $K^{*}$--$\nu_h^{*}$ Trade-off and Three Operating Regimes}\label{sec:dyn_xlayer}

We extend (P2) by promoting the handover rate $\nu_h$ to a continuous decision variable bounded above by a hardware-imposed handover-capability ceiling $\nu_h^{\max}$ (typically $2$--$10\,\mathrm{s}^{-1}$ for 3GPP NR handover, $\sim\!100\,\mathrm{s}^{-1}$ for fast digital beam re-pointing, and $\infty$ for the idealised continuous-NN limit of Remark~5). The per-target effective resource consumption per BS becomes
\begin{equation}
    \rho_{\mathrm{eff}}(\nu_h) \;=\; \rho_0\,\big(1+\nu_h\tau_{\mathrm{ho}}\big),
    \label{eq:rho_eff_dyn}
\end{equation}
where $\tau_{\mathrm{ho}}\!\in\![10,100]\,\mathrm{ms}$ aggregates the per-handover signalling, beam re-pointing, and CU re-initialisation cost (Sec.~\ref{sec:dynamic_cluster}). The dynamic cross-layer problem is
\begin{align}
    \textbf{(P2-DYN)}\;\;\max_{\rho_0,\theta_m,K,\nu_h}\;&\widehat C_{\mathrm{comm}}(\rho_0,\theta_m;\nu_h) \nonumber\\
    \mathrm{s.t.}\;\;& \mathbb{E}[\rv{T}_{\mathrm{loss}}^{\mathrm{dyn}}(K,\nu_h)]\!\ge\!\tau_{\mathrm{req}}, \label{eq:p2dyn_c1}\\
    & P_{\mathrm{block}}\!\big(\rho_{\mathrm{eff}}(\nu_h),K\big)\!\le\!\epsilon_{\mathrm{macro}}, \label{eq:p2dyn_c2}\\
    & (1-p_{\mathrm{link}})^{K}\!\le\!\epsilon_{\mathrm{rel}}, \label{eq:p2dyn_c_rel}\\
    & 0\!\le\!\nu_h\!\le\!\nu_h^{\max}, \label{eq:p2dyn_c3}\\
    & \theta_{\min}\!\le\!\theta_m\!\le\!\min\{\theta_{\max},\theta_{\mathrm{crit}}\}. \nonumber
\end{align}
The kinematic constraint \eqref{eq:p2dyn_c1} is the dynamic counterpart of \eqref{eq:p2_c1}, with $\rv{T}_{\mathrm{loss}}^{\mathrm{coop}}$ replaced by the Theorem~1-DYN expression \eqref{eq:thm1_dyn}. The blocking constraint \eqref{eq:p2dyn_c2} substitutes $\rho_{\mathrm{eff}}$ for $\rho_0$, so handover overhead enters the Poisson load mean $K\lambda_T\rho_{\mathrm{eff}}/\lambda_b$ on equal footing with the sensing duty cycle.

The new constraint \eqref{eq:p2dyn_c_rel} is the \textit{cluster-wide link-reliability constraint}: let $p_{\mathrm{link}}\!\in\!(0,1]$ denote the per-link availability (probability that any individual cooperator's link is usable at any given time, accounting for deep blockage, beam misalignment, hardware faults, and interference spikes; typical 6G values $0.85$--$0.99$ depending on UMa / UMi / RMa scenario), and let $\epsilon_{\mathrm{rel}}\!\in\![10^{-3},10^{-2}]$ denote the operator's tolerance for simultaneous failure of all $K$ cooperators. Under the standard independent-failure assumption across spatially separated cooperators (justified by the geometric dispersion $d_{\min}^{(b)}\!=\!50$ m of K-NN BSs and the frequency-selective fading independence across BSs), the simultaneous-failure probability is $(1-p_{\mathrm{link}})^{K}$, and \eqref{eq:p2dyn_c_rel} inverts to a hard lower bound on the cluster size,
\begin{equation}
    K\;\ge\;K_{\mathrm{rel}}(p_{\mathrm{link}},\epsilon_{\mathrm{rel}})\;\triangleq\;\left\lceil\,\frac{\ln\epsilon_{\mathrm{rel}}}{\ln(1-p_{\mathrm{link}})}\,\right\rceil,
    \label{eq:K_rel}
\end{equation}
typically $K_{\mathrm{rel}}\!\in\!\{2,3,4\}$ for representative 6G channel models (e.g., $p_{\mathrm{link}}\!=\!0.90$, $\epsilon_{\mathrm{rel}}\!=\!10^{-3}$ gives $K_{\mathrm{rel}}\!=\!3$). The reliability constraint is orthogonal to the kinematic, blocking, and handover constraints and acts purely as a hard floor on $K$.

\textbf{Theorem 1$'$-DYN (Dynamic Cross-Layer Optimum):} \textit{Let $K^{*}_{\mathrm{static}}\!=\!\lceil 4\pi\eta\lambda_b D/(\mu_{\mathrm{sj}}\epsilon_{\mathrm{micro}})\rceil$ be the static cross-layer optimum of Theorem~1$'$ and let}
\begin{equation}
    K_{\min}(\nu_h^{\max}) \;\approx\; \frac{\pi\lambda_b D\,\ln^{2}\!\big(K^{*}_{\mathrm{static}}\big)}{\nu_h^{\max}}
    \label{eq:Kmin_dyn}
\end{equation}
\textit{denote the smallest cluster size consistent with a handover-rate ceiling $\nu_h^{\max}$ (obtained by inverting Corollary~5 at $K\!=\!K_{\min}$). The optimum of {(P2-DYN)} in $(K,\nu_h)$ falls in one of three operating regimes:}
\begin{equation}
\!\!(K^{*},\,\nu_h^{*}) =
    \begin{cases}
        \!\big(\max\{1,K_{\mathrm{rel}}\},\,\nu_h^{\min}(K^{*})\big), & \!\text{(i)}\\[2pt]
        \!\big(\max\{K_{\min}(\nu_h^{\max}),K_{\mathrm{rel}}\},\,\nu_h^{\max}\big), & \!\text{(ii)}\\[2pt]
        \!\big(\max\{K^{*}_{\mathrm{static}},K_{\mathrm{rel}}\},\,0\big), & \!\text{(iii)}
    \end{cases}
    \label{eq:three_regimes}
\end{equation}
\textit{where the three regimes are defined by the hardware handover-rate ceiling $\nu_h^{\max}$ relative to the natural scales of the problem:}
\begin{align*}
    \text{(i) Handover-rich:}\quad & \nu_h^{\max}\!\ge\!\nu_h^{\min}(1),\\
    \text{(ii) Mid-budget:}\quad & \pi\lambda_b D/K^{*}_{\mathrm{static}}\!<\!\nu_h^{\max}\!<\!\nu_h^{\min}(1),\\
    \text{(iii) Handover-starved:}\quad & \nu_h^{\max}\!\to\!0,
\end{align*}
\textit{where $\nu_h^{\min}(1)\!=\!\pi\lambda_b D\,\ln^{2}(K^{*}_{\mathrm{static}})$ is the handover rate that allows the single-BS cluster to meet the kinematic constraint, and $\pi\lambda_b D/K^{*}_{\mathrm{static}}$ is the Bessel small/large-argument crossover rate of \eqref{eq:thm1_dyn_asympt} evaluated at $K\!=\!K^{*}_{\mathrm{static}}$, marking the onset of the exponential-lift regime where the doubly-logarithmic approximation of Corollary~5 takes hold. Regime (iii) recovers the static Theorem~1$'$ exactly; regime (i) is the unconstrained dynamic optimum in which the cluster collapses to a single BS and all kinematic burden is carried by handover.}

\textit{Proof sketch.} On the active kinematic-constraint surface Corollary~5 gives $\nu_h^{\min}(K)$; substituting into the per-target cost yields $\mathcal{J}(K)\!=\!K\rho_0\!+\!K\nu_h^{\min}(K)\tau_{\mathrm{ho}}\rho_0$. Using $K\nu_h^{\min}(K)\!=\!\pi\lambda_b D\ln^{2}(K^{*}_{\mathrm{static}}/K)$ collapses the handover term, giving $\mathcal{J}(K)\!=\!K\rho_0+\pi\lambda_b D\tau_{\mathrm{ho}}\rho_0\ln^{2}(K^{*}_{\mathrm{static}}/K)$. The second term is doubly logarithmic in $K$, so approximately $K$-independent, so $\partial\mathcal{J}/\partial K\!\approx\!\rho_0\!>\!0$ and the unconstrained minimum sits at $K\!=\!1$. The constraint $\nu_h\!\le\!\nu_h^{\max}$ binds precisely when $\nu_h^{\min}(K\!=\!1)\!>\!\nu_h^{\max}$, giving the three-regime case split. $\hfill\blacksquare$

\textit{Discussion.} The static kinematic optimum $K^{*}_{\mathrm{static}}(\lambda_b)\!=\!4\pi\eta\lambda_b D/(\mu_{\mathrm{sj}}\epsilon_{\mathrm{micro}})$ scales linearly in $\lambda_b$. For typical macro ($\lambda_b\!=\!10\,\mathrm{BS/km}^{2}$, $\eta\!=\!0.3$, $D\!=\!1\,\mathrm{m}^{2}/\mathrm{s}$, $\epsilon_{\mathrm{micro}}\!=\!10^{-2}$, $\mu_{\mathrm{sj}}\!=\!10^{-2}\,\mathrm{s}^{-1}$), $K^{*}_{\mathrm{static}}\!\approx\!0.4$, so single-BS sensing already meets the kinematic constraint and the operational cluster size is fixed by the reliability floor $K_{\mathrm{rel}}\!\in\!\{2,3,4\}$ alone. Beyond $\lambda_b\!\gtrsim\!80\,\mathrm{BS/km}^{2}$ the static $K^{*}$ overtakes $K_{\mathrm{rel}}$ and inflates linearly, reaching $\sim\!40$ at the realistic very-dense-small-cell ceiling $\lambda_b\!\sim\!10^{3}\,\mathrm{BS/km}^{2}$. The dynamic framework pins $K^{*}$ at $K_{\mathrm{rel}}$ for all densities, off-loading kinematic survival onto a doubly-logarithmic handover rate (Corollary~5). The value of the dynamic upgrade is density-dependent: at typical macro the two frameworks coincide, and at $\lambda_b\!\sim\!10^{3}\,\mathrm{BS/km}^{2}$ the static recommendation requires roughly $13\times$ the dynamic per-target backhaul commitment at handover overhead $\nu_h^{\min}\!\sim\!10^{-2}\,\mathrm{s}^{-1}$ (within 3GPP NR capability).

\textbf{Remark 8 (Independence Assumption in the Reliability Constraint).} \textit{The cluster-wide failure probability $(1-p_{\mathrm{link}})^{K}$ assumes link failures are independent across the $K$ cooperators. Two correlated-failure mechanisms can violate this in practice: shared blockers (e.g., a single large vehicle blocking the LoS path to multiple BSs simultaneously) and correlated atmospheric fading. Both push the effective failure probability above the independent baseline, requiring slightly larger $K_{\mathrm{rel}}$ in practice. The independent-failure baseline of \eqref{eq:K_rel} is therefore a lower bound on the engineering cluster size, consistent with the conservative-upper-bound philosophy of the rest of the framework.}

\textbf{Theorem 3-DYN (Lifted Capacity Ceiling under Dynamic Clustering):} \textit{Substituting the Theorem~1$'$-DYN regime-(i) optimum $K^{*}\!=\!\max\{1,K_{\mathrm{rel}}\}\!=\!K_{\mathrm{rel}}$ and $\nu_h^{*}\!=\!\nu_h^{\min}(K_{\mathrm{rel}})$ from Corollary~5 into the per-BS mean-load constraint $K^{*}\rho_{\mathrm{eff}}(\nu_h^{*})\lambda_T/\lambda_b\!\le\!\rho_{\max}$, and noting that $K^{*}\nu_h^{*}\tau_{\mathrm{ho}}\!=\!\pi\lambda_b D\tau_{\mathrm{ho}}\ln^{2}(K^{*}_{\mathrm{static}}(\lambda_b)/K_{\mathrm{rel}})$, the dynamic-cluster sensing-capacity ceiling is}
\begin{equation}
\lambda_T^{\max,\mathrm{dyn}}(\lambda_b) \!=\! \dfrac{\rho_{\max}\lambda_b}{\rho_0\!\left[K_{\mathrm{rel}}\!+\!\pi\lambda_b D\tau_{\mathrm{ho}}\ln^{2}\!\!\big(\tfrac{K^{*}_{\mathrm{static}}(\lambda_b)}{K_{\mathrm{rel}}}\big)\right]}
\label{eq:capacity_ceiling_dyn}
\end{equation}
\textit{where $K^{*}_{\mathrm{static}}(\lambda_b)\!=\!4\pi\eta\lambda_b D/(\mu_{\mathrm{sj}}\epsilon_{\mathrm{micro}})$ is the kinematic optimum of Theorem~1$'$, linear in $\lambda_b$. When $K^{*}_{\mathrm{static}}(\lambda_b)\!\le\!K_{\mathrm{rel}}$ (sparse densities) the kinematic constraint is auto-met at $K\!=\!K_{\mathrm{rel}}$ with no handover ($\nu_h^{*}\!=\!0$) and the overhead term vanishes. For all realistic 6G densities the overhead term remains $<\!1\%$ of $K_{\mathrm{rel}}$ and the ceiling reduces to $\lambda_T^{\max,\mathrm{dyn}}\!\approx\!\rho_{\max}\lambda_b/(K_{\mathrm{rel}}\rho_0)$: classical density-scaling sensing capacity holds up to a constant reliability divisor $K_{\mathrm{rel}}$. The asymptotic crossover density at which the overhead term would dominate,}
\begin{equation}
    \lambda_b^{c,\mathrm{dyn}}:\quad K_{\mathrm{rel}}\;=\;\pi\lambda_b^{c,\mathrm{dyn}} D\tau_{\mathrm{ho}}\,\ln^{2}\!\big(K^{*}_{\mathrm{static}}(\lambda_b^{c,\mathrm{dyn}})/K_{\mathrm{rel}}\big),
    \label{eq:lambda_b_c_dyn}
\end{equation}
\textit{evaluates to $\sim\!4\!\times\!10^{5}\,\mathrm{BS/km}^{2}$ ($\sim\!1$ BS per $2.5\,\mathrm{m}^{2}$) --- a physically meaningless density --- so the dynamic ceiling effectively never saturates.}

\textit{Numerical magnitude.} For typical 6G parameters ($\tau_{\mathrm{ho}}\!=\!50\,\mathrm{ms}$, $D\!=\!1\,\mathrm{m}^{2}/\mathrm{s}$, $\eta\!=\!0.3$, $\mu_{\mathrm{sj}}\!=\!10^{-2}\,\mathrm{s}^{-1}$, $\epsilon_{\mathrm{micro}}\!=\!10^{-2}$, $K_{\mathrm{rel}}\!=\!3$): the static kinematic optimum $K^{*}_{\mathrm{static}}(\lambda_b)$ first exceeds the reliability floor at $\lambda_b^{c,\mathrm{static}}\!\approx\!80\,\mathrm{BS/km}^{2}$ and grows to $\sim\!40$ at the realistic very-dense-small-cell ceiling $\lambda_b\!\sim\!10^{3}\,\mathrm{BS/km}^{2}$. Beyond this density 6G is not deploying any meaningful infrastructure. The static Theorem~3 ceiling plateaus at $\sim\!265\,\mathrm{targets/km}^{2}$ above $\lambda_b\!\approx\!80\,\mathrm{BS/km}^{2}$, while the dynamic ceiling continues linearly: the capacity gap is $\sim\!1\!\times$ at typical macro, $\sim\!1.3\!\times$ at $\lambda_b\!=\!100$, $\sim\!6\!\times$ at small-cell $\lambda_b\!=\!500$, and $\sim\!12\!\times$ at the realistic densification ceiling $\lambda_b\!=\!10^{3}\,\mathrm{BS/km}^{2}$. Equivalently, an operator densifying from $\lambda_b\!=\!10$ to $10^{3}\,\mathrm{BS/km}^{2}$ sees the static-recommended cluster size inflate from $K\!=\!K_{\mathrm{rel}}$ to $K\!\approx\!40$, while the dynamic framework holds $K\!=\!K_{\mathrm{rel}}$ throughout: a $\sim\!13\times$ reduction in per-target backhaul commitment at the densification ceiling.

\textbf{Remark 6 (Static Results as the $\nu_h\!\to\!0$ Corner).} \textit{The static cross-layer Theorem~1$'$ and the phase-conditioned ceiling Theorem~3 of the preceding subsection are recovered exactly by Theorem~1$'$-DYN regime~(iii) with $\nu_h^{\max}\!=\!0$ and by Theorem~3-DYN in the asymptotic ultra-dense regime $\lambda_b\!\gg\!\lambda_b^{c,\mathrm{dyn}}$, respectively. The original static results are therefore not overturned but embedded as a corner case of the richer dynamic framework. The ``$\lambda_b$-independence ceiling'' of the static framework does not bind in the dynamic case: the crossover \eqref{eq:lambda_b_c_dyn} is pushed beyond $\sim\!10^{5}\,\mathrm{BS/km}^{2}$, so the dynamic ceiling exhibits classical density-scaling (modulo $K_{\mathrm{rel}}$) throughout the practical operating range, while the static ceiling already caps at $\lambda_b\!\sim\!80\,\mathrm{BS/km}^{2}$.}

\textbf{Remark 7 (Binding Design Lever: Per-Link Availability $p_{\mathrm{link}}$).} \textit{Throughout all realistic 6G operating densities ($\lambda_b\!\le\!10^{3}\,\mathrm{BS/km}^{2}$), the dynamic ceiling \eqref{eq:capacity_ceiling_dyn} is dominated by the $K_{\mathrm{rel}}$ term in the denominator while the handover-overhead $\pi\lambda_b D\tau_{\mathrm{ho}}\ln^{2}(\cdot)$ remains $<\!1\%$ of $K_{\mathrm{rel}}$. The binding design lever for sensing capacity is therefore the per-link availability $p_{\mathrm{link}}$ (which sets $K_{\mathrm{rel}}$), driven by channel-quality engineering, beam-management robustness, and blockage mitigation. The handover signalling latency $\tau_{\mathrm{ho}}$ enters only as a sub-leading correction unless the network is densified to the academic regime $\lambda_b\!\gtrsim\!10^{5}\,\mathrm{BS/km}^{2}$.}

\subsection{Cross-Layer Solution Structure}\label{sec:closed_form}
We reduce the probabilistic kinematic constraint \eqref{eq:p2_c1} to a deterministic surrogate via the exponential-tail FPT approximation $\mathbb{P}(\rv{T}_{\text{loss}}^{\text{coop}}\!\ge\!T_{\text{req}})\!\approx\!\exp(-T_{\text{req}}/\mathbb{E}[\rv{T}_{\text{loss}}^{\text{coop}}])$ \cite{redner2001guide}, which inverts to the MTLT lower bound
\begin{equation}
    \mathbb{E}[\rv{T}_{\text{loss}}^{\text{coop}}]\ge\tau_{\text{req}}\triangleq\frac{-T_{\text{req}}}{\ln(1\!-\!\epsilon_{\text{micro}})}\xrightarrow{\epsilon_{\text{micro}}\ll 1}\frac{\eta}{\mu_{\text{sj}}\epsilon_{\text{micro}}}.
    \label{eq:tau_req}
\end{equation}
The $1/\epsilon_{\text{micro}}$ inflation is the exponential-tail cost of strict reliability. Because Theorem~1's super-critical MTLT $K/(4\pi\lambda_b D)$ is independent of $(\rho_0,\theta_m)$, the kinematic constraint is encoded discretely through $K^{*}$ and the cross-layer optimum decouples into three sequential stages: (i) $K^{*}$ from \eqref{eq:Kstar_xlayer}; (ii) $\theta_m^{*}\!=\!\min\{\theta_{\max},\theta_{\mathrm{crit}},\arg\max\bar R(\theta_m)\}$ (where $\bar R(\theta_m)\triangleq\mathbb{E}[\log_2(1+\mathrm{SINR}_{\text{UE}}(\theta_m))]$ is the typical-UE ergodic rate) from rate-maximisation clipped at the beamwidth trap; (iii) $\rho_0^{*}$ from inverting $P_{\text{block}}(\rho_0,K^{*})\!=\!\epsilon_{\text{macro}}$. Sweeping the QoS coverage fraction $\eta$ traces the Pareto frontier between sensing robustness and communication capacity.

\section{Numerical Results}
\label{sec:numerical_results}

We validate the framework on a $10\!\times\!10\,\mathrm{km}^2$ network with baseline macro density $\lambda_b\!=\!10\,\mathrm{BS/km^2}$ (3GPP UMa) and target diffusion $D\!=\!1\,\mathrm{m^2/s}$ (low-altitude urban UAV). The full parameter set is in Table~\ref{tab:parameters}. Each data point averages $2\!\times\!10^5$ Brownian trajectories over 200 BS realisations of a 2D HPPP with hard-core constraint $d_{\min}^{(b)}\!=\!50\,\mathrm{m}$.

\begin{table}[t]
\centering
\caption{System Parameters for Simulation}
\label{tab:parameters}
\renewcommand{\arraystretch}{1.10}
\setlength{\tabcolsep}{5pt}
\begin{tabular*}{\columnwidth}{@{\extracolsep{\fill}}l l@{}}
\toprule
\textbf{Parameter} & \textbf{Value} \\
\midrule
\multicolumn{2}{@{}l}{Network geometry and target kinematics}\\
BS density $\lambda_b$               & $10\,\mathrm{BS/km^{2}}$ \\
Diffusion coefficient $D$            & $1\,\mathrm{m^{2}/s}$ \\
Mean sojourn time $1/\mu_{\mathrm{sj}}$ & $100\,\mathrm{s}$ \\
Min.\ BS spacing $d_{\min}^{(b)}$    & $50\,\mathrm{m}$ \\
Target near-field guard $r_{\min}$   & $10\,\mathrm{m}$ \\
\midrule
\multicolumn{2}{@{}l}{QoS and resource budget}\\
QoS coverage fraction $\eta$         & $0.3$ \\
Kinematic outage $\epsilon_{\mathrm{micro}}$ & $10^{-2}$ \\
Macro blocking $\epsilon_{\mathrm{macro}}$ & $0.05$ \\
Link availability $p_{\mathrm{link}}$, tolerance $\epsilon_{\mathrm{rel}}$ & $0.9,\ 10^{-3}$ \\
Per-cooperator $\rho_0$, per-BS budget $\rho_{\max}$ & $0.01,\ 0.10$ \\
Reliability floor $K_{\mathrm{rel}}$ & $3$ \\
Handover signalling $\tau_{\mathrm{ho}}$ & $50\,\mathrm{ms}$ \\
\midrule
\multicolumn{2}{@{}l}{Radio and propagation}\\
Radar detection threshold $\gamma_s$ & $5\,\mathrm{dB}$ \\
BS transmit power $P_t$              & $46\,\mathrm{dBm}$ \\
Target RCS $\sigma$                  & $1\,\mathrm{m^{2}}$ \\
Antenna main-lobe gain $G_m$         & $30\,\mathrm{dBi}$ \\
Side-lobe ratio $\zeta$              & $-40\,\mathrm{dB}$ \\
LoS blockage rate $\beta$            & $10^{-2}\,\mathrm{m^{-1}}$ \\
Clutter path-loss exponent $\alpha_c$ & $3.5$ \\
\bottomrule
\end{tabular*}
\end{table}

\begin{figure}[!t]
\centering
\subfigure[Single-BS MTLT vs $\lambda_b$]{\includegraphics[width=0.48\columnwidth]{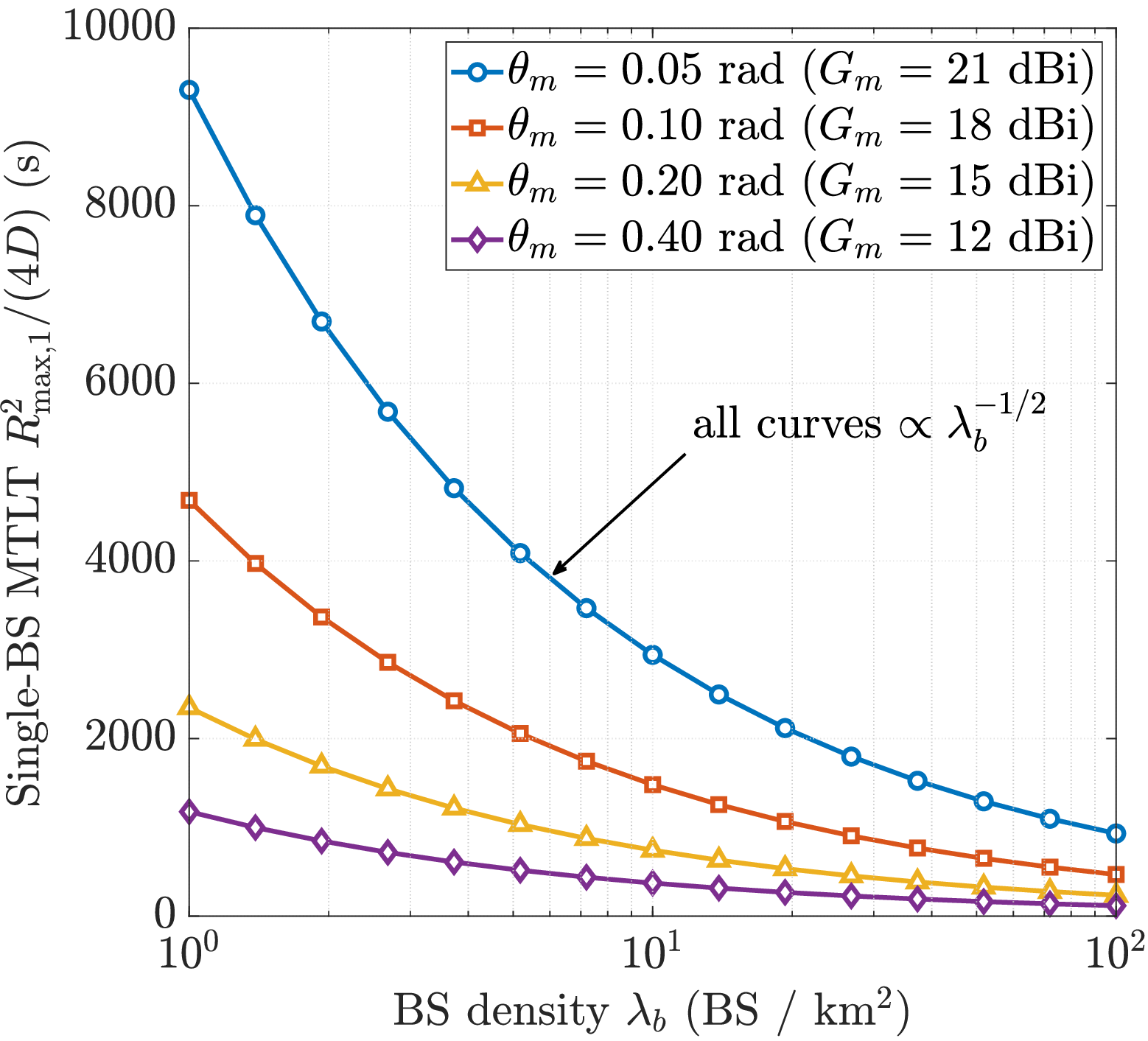}}\hfill
\subfigure[Cooperative MTLT vs $K$]{\includegraphics[width=0.48\columnwidth]{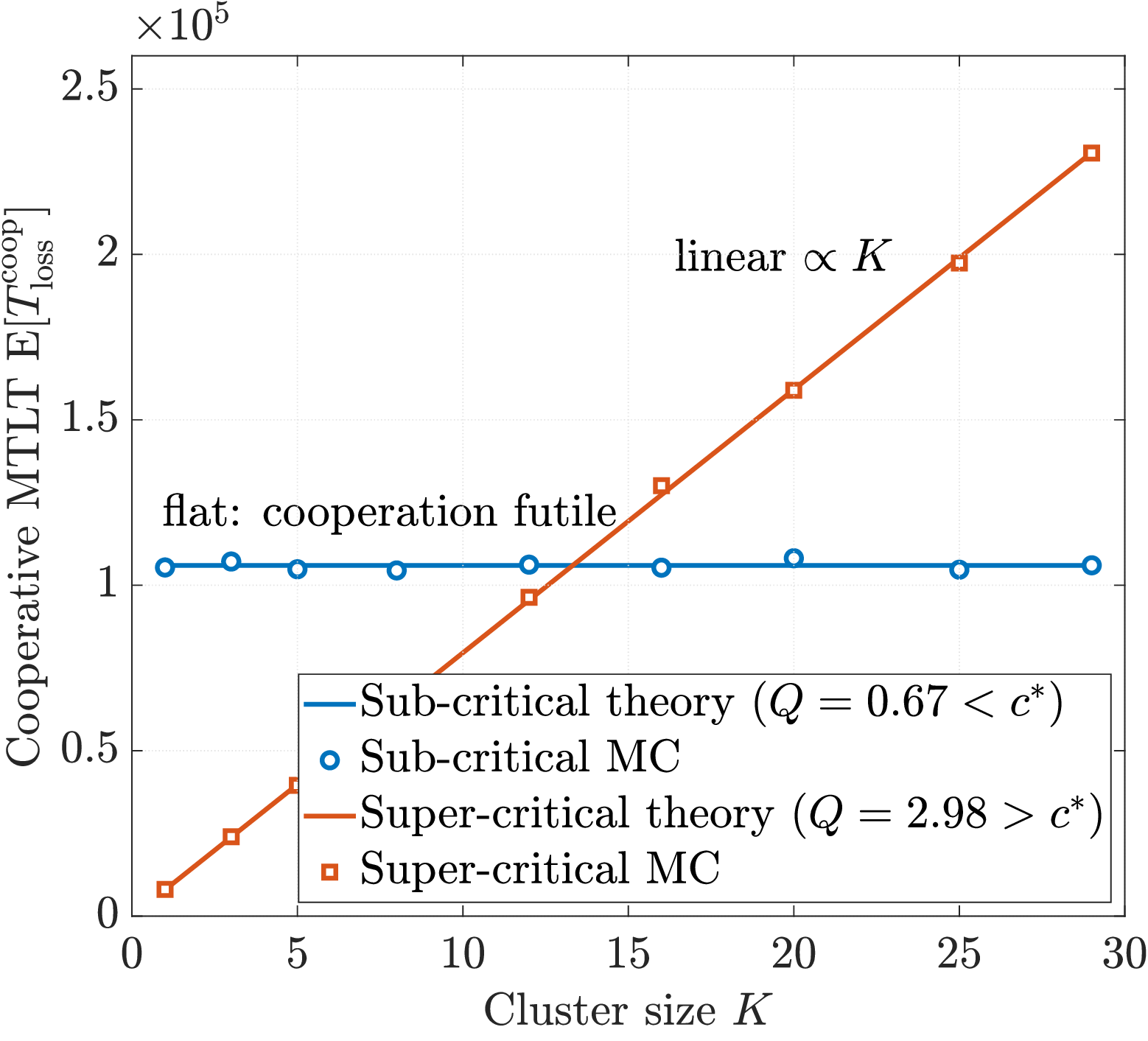}}
\caption{(a) Single-BS MTLT $R_{\max,1}^{2}/(4D)$ decays as $\lambda_b^{-1/2}$ for all $\theta_m$, confirming Remark~1: antenna sharpening cannot mitigate densification collapse because $\bar G\!\approx\!1$. (b) Cooperative MTLT validates Theorem~1: sub-critical ($Q\!<\!c^{*}$) is $K$-independent; super-critical ($Q\!>\!c^{*}$) is linear-in-$K$ with slope $1/(4\pi\lambda_b D)$.}
\label{fig:single_and_phase}
\end{figure}

\begin{figure}[!t]
\centering
\subfigure[Heavy-tail CCDF at criticality]{\includegraphics[width=0.48\columnwidth]{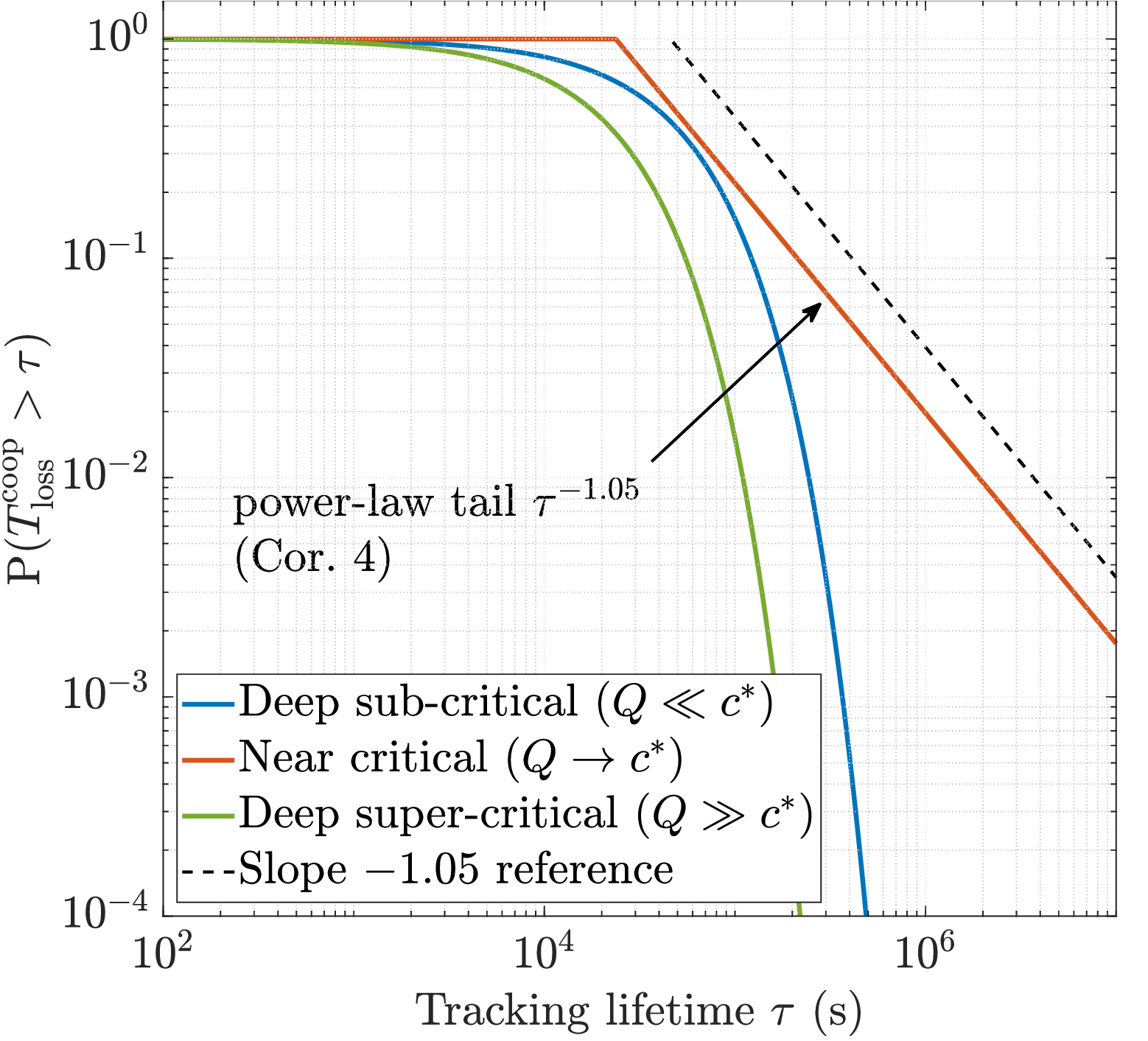}}\hfill
\subfigure[Beamwidth trap]{\includegraphics[width=0.48\columnwidth]{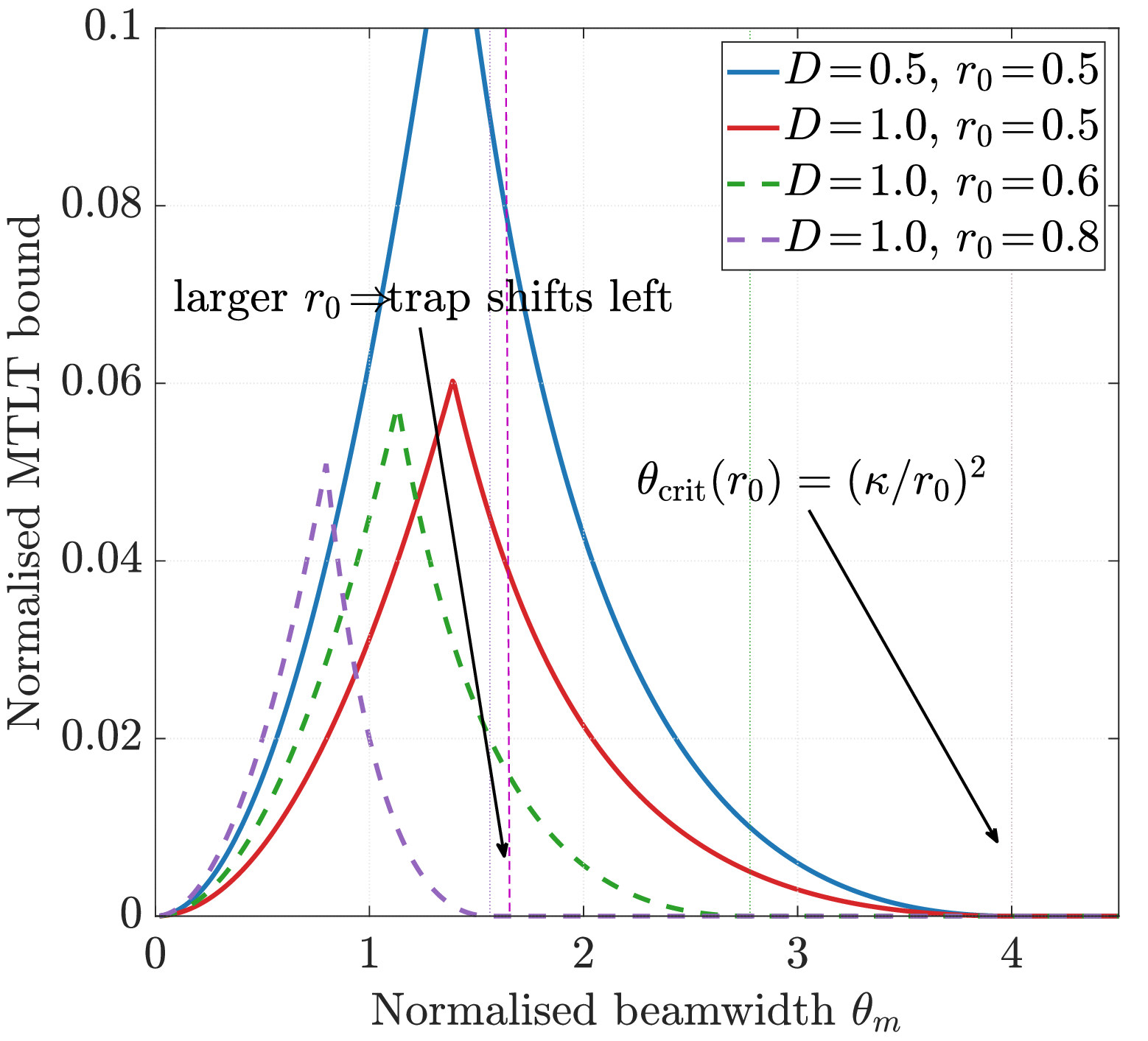}}
\caption{(a) Near percolation criticality the MTLT distribution acquires the universal tail $\mathbb{P}(\rv{T}_{\text{loss}}\!>\!\tau)\!\sim\!\tau^{-1.05}$ (parallel to the dashed reference), confirming Corollary~4. (b) Beamwidth trap (Theorem~2): MTLT rises with $\theta_m$ in the angular-escape regime and collapses past $\theta_{\text{crit}}\!=\!(\kappa/r_0)^{2}$ where radial escape dominates; larger $r_0$ tightens the trap.}
\label{fig:heavytail_and_beamtrap}
\end{figure}

\begin{figure}[!t]
\centering
\subfigure[Phase-conditioned capacity ceiling]{\includegraphics[width=0.48\columnwidth]{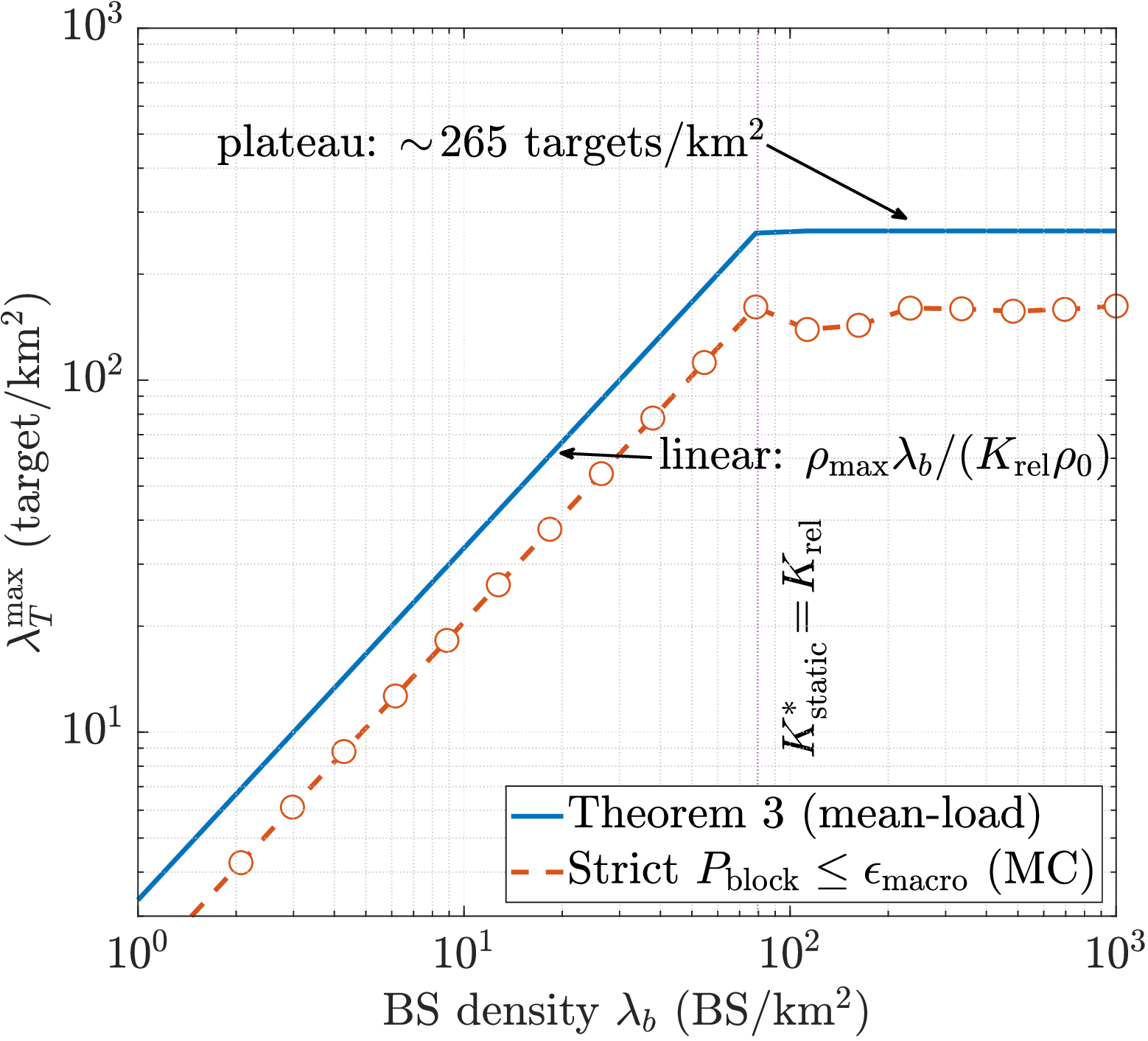}}\hfill
\subfigure[Pareto frontier (static baseline)]{\includegraphics[width=0.48\columnwidth]{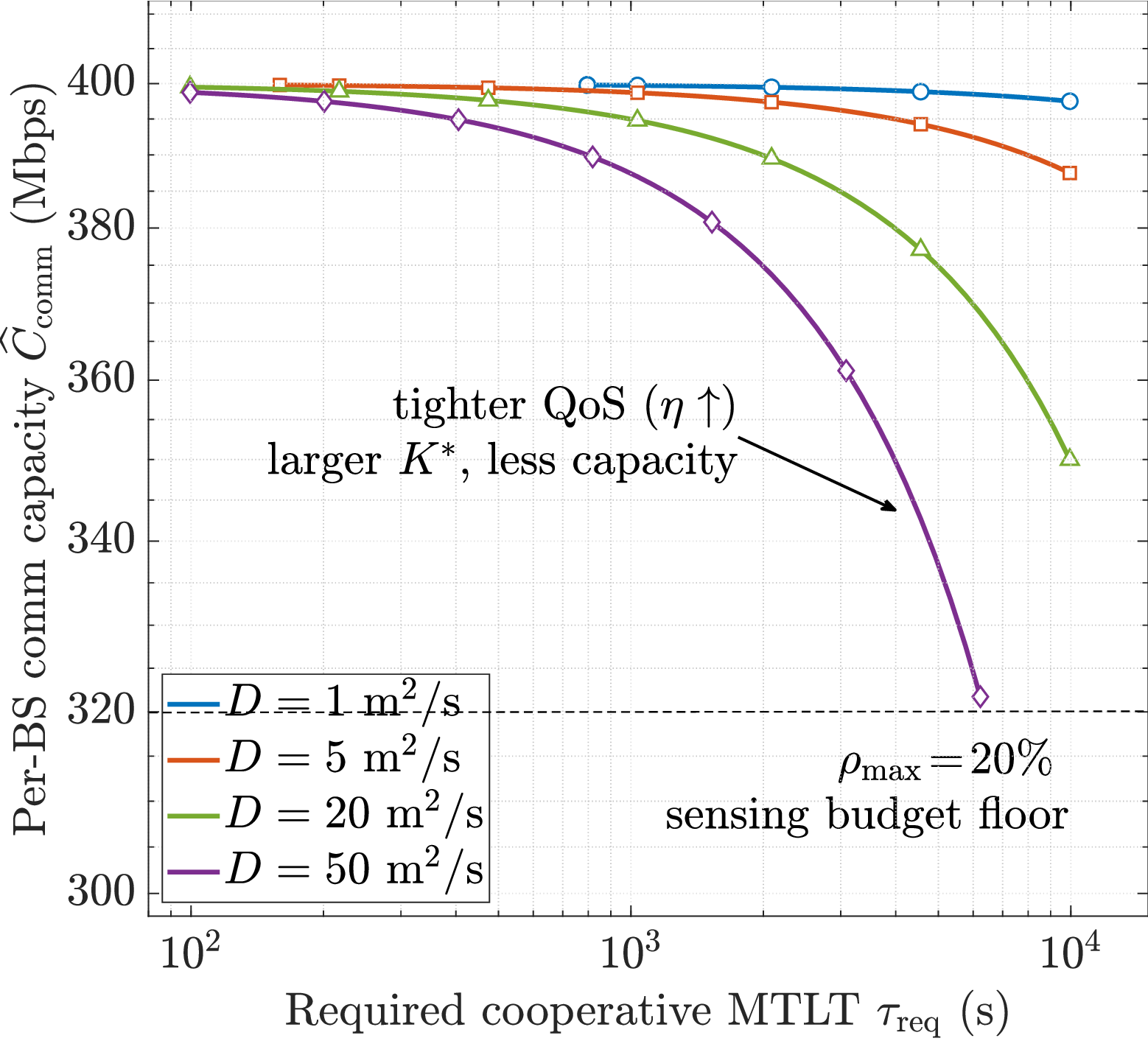}}
\caption{(a) Static Theorem~3 capacity ceiling: linear sub-critical ($\rho_{\max}\lambda_b/\rho_0$) transitions to $\lambda_b$-independent plateau super-critical; Monte-Carlo with strict $\epsilon_{\text{macro}}$ sits a constant $c_{\epsilon_{\text{macro}}}\!\approx\!0.37$ below the mean-load bound. (b) Cross-layer Pareto frontier between $\widehat C_{\text{comm}}$ and $\tau_{\text{req}}$ for four diffusion regimes (Theorem~1$'$), evaluated at $\lambda_b\!=\!10^{2}$ BS/km$^{2}$, $\lambda_T\!=\!10$ targets/km$^{2}$, $\rho_0\!=\!0.005$, $\rho_{\max}\!=\!0.20$ (denser-than-baseline setting chosen so all four $D$-regimes operate super-critically); kinematic agility $D$ dominates the slope.}
\label{fig:capacity_and_pareto}
\end{figure}

We begin with the static framework in Figs.~\ref{fig:single_and_phase}--\ref{fig:capacity_and_pareto}. In Fig.~\ref{fig:single_and_phase}(a), four representative beamwidths are swept across three decades of $\lambda_b$; it can be observed that all four curves decay as $\lambda_b^{-1/2}$ irrespective of $\theta_m$ --- a visual confirmation that antenna sharpening does not arrest the densification collapse (Remark~1). Panel~(b) overlays Monte-Carlo runs on either side of the percolation threshold ($\lambda_b\!=\!0.5$ and $10\,\mathrm{BS/km^{2}}$, giving $Q\!=\!0.67$ and $2.98$), and the regime contrast is sharp: $K$-independent below $c^{*}$, linear in $K$ above. Near criticality, the $10^{-2}$ quantile of the MTLT distribution runs roughly an order of magnitude below its mean (Fig.~\ref{fig:heavytail_and_beamtrap}(a)); any QoS framework keyed on $\mathbb{E}[\rv{T}_{\text{loss}}]$ thus systematically under-provisions the worst $1\%$ of trajectories, motivating the percentile-graded constraint of \eqref{eq:p2_c1}. Panel~(b) overlays four $(D,r_0)$ scenarios; the leftward shift of $\theta_{\mathrm{crit}}$ with $r_0$ rules out a single global beamwidth cap. The headline static result appears in Fig.~\ref{fig:capacity_and_pareto}(a): $\lambda_T^{\max}$ scales linearly up to the reliability crossover $\lambda_b^{\mathrm{c}}\!\approx\!80\,\mathrm{BS/km^{2}}$ and saturates at $\sim\!265\,\mathrm{targets/km^{2}}$ thereafter. Panel~(b) sweeps $\eta\!\in\![0.01,1]$ across four diffusion regimes $D\!\in\!\{1,5,20,50\}\,\mathrm{m^{2}/s}$, where kinematic agility $D$ dominates the Pareto slope --- a $D\!=\!20$ UAV costs roughly an order of magnitude more capacity per second of $\tau_{\mathrm{req}}$ than a $D\!=\!1$ ground target.

\begin{figure}[!t]
\centering
\subfigure[Dynamic MTLT vs $\nu_h$]{\includegraphics[width=0.48\columnwidth]{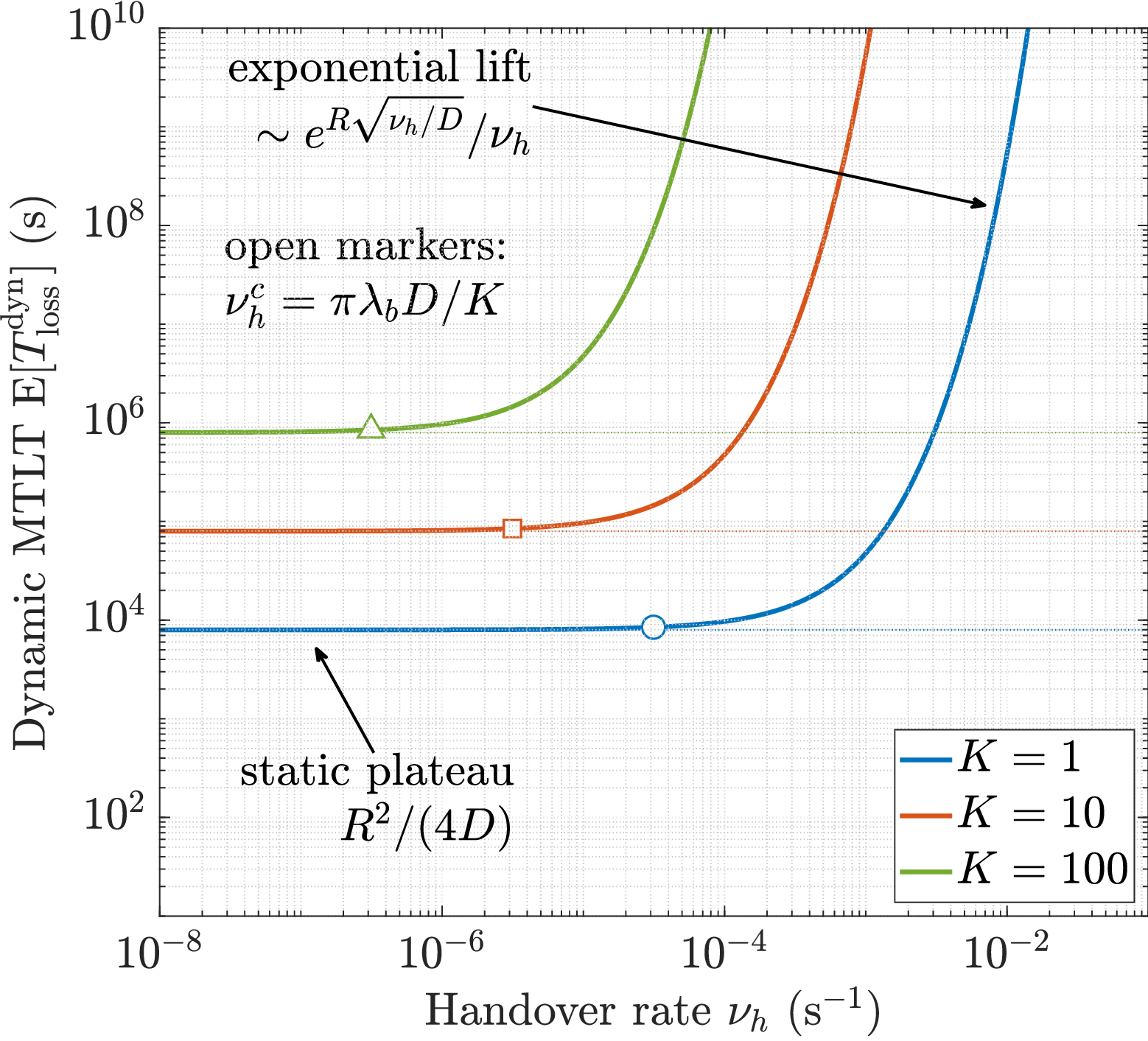}}\hfill
\subfigure[Lifted capacity ceiling]{\includegraphics[width=0.48\columnwidth]{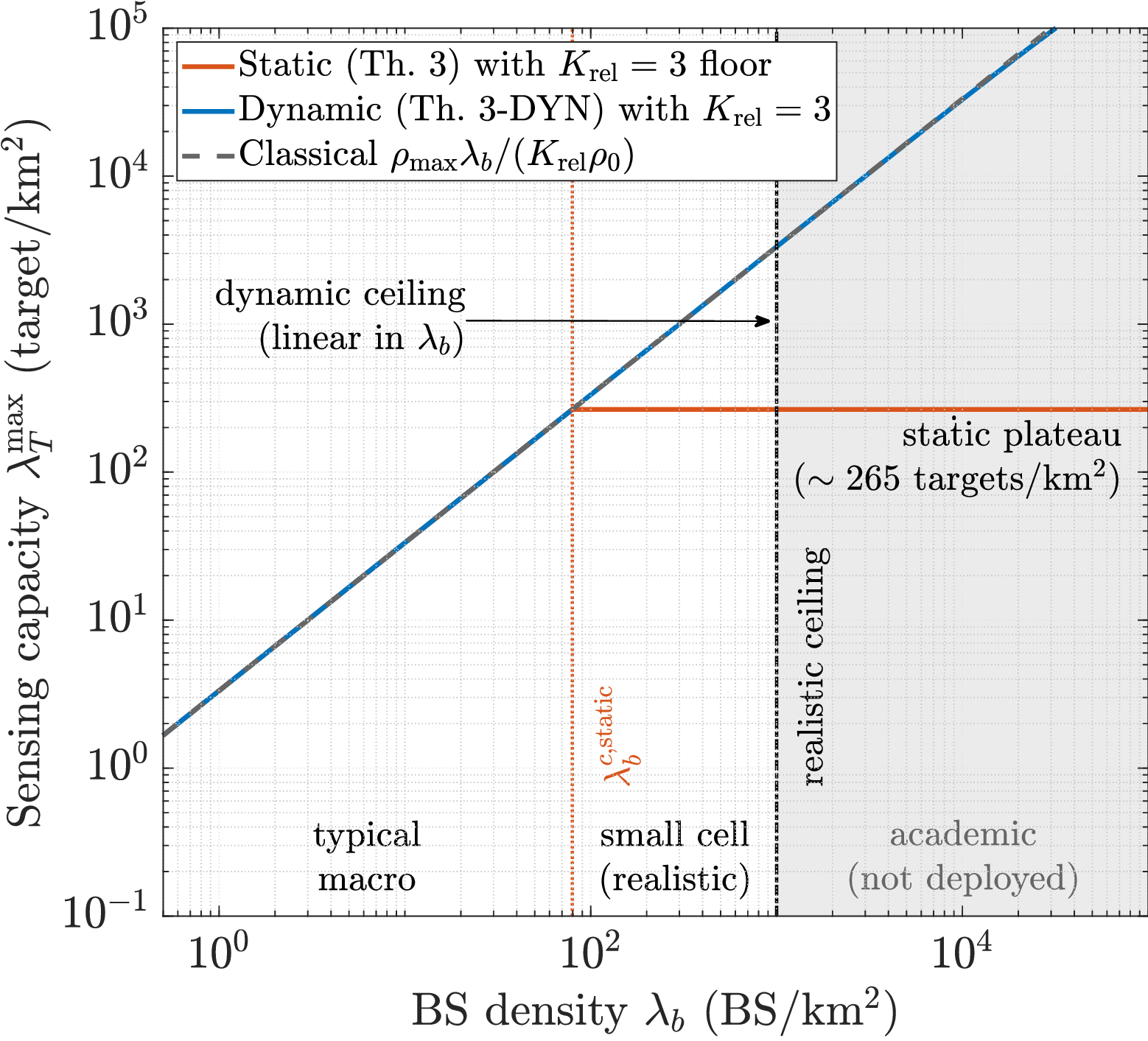}}
\caption{(a) Theorem~1-DYN: the Bessel-$I_0$ MTLT has a slow-handover plateau at the static $R^{2}/(4D)$ and a super-polynomial blow-up beyond $\nu_h^{c}\!=\!\pi\lambda_b D/K$ (open markers), validating Corollary~5's doubly-logarithmic scaling. (b) Theorem~3-DYN: static (red) plateaus at $\sim\!265$ targets/km$^{2}$ above $\lambda_b\!\approx\!80$ BS/km$^{2}$; dynamic (blue) preserves linear $\rho_{\max}\lambda_b/(K_{\mathrm{rel}}\rho_0)$ scaling throughout the realistic range, opening a $\sim\!12\times$ gap at the densification ceiling $\lambda_b\!=\!10^{3}$ BS/km$^{2}$.}
\label{fig:dyn_results}
\end{figure}

Fig.~\ref{fig:dyn_results} demonstrates the dynamic-cluster contribution. Panel~(a) sweeps $\nu_h$ across eight decades for $K\!\in\!\{1,10,100\}$; Corollary~5's doubly-logarithmic scaling is operationally cheap --- a $10\times$ MTLT extension requires $\nu_h$ to grow by only $(\ln 10)^{2}\!\approx\!5.3$, well within 3GPP NR handover capability. Panel~(b) quantifies the dividend at the small-cell ceiling $\lambda_b\!=\!10^{3}\,\mathrm{BS/km^{2}}$: a $\sim\!12\!\times$ capacity lift over the static plateau and a $\sim\!13\!\times$ per-target cluster-commitment reduction. Together, Figs.~\ref{fig:single_and_phase}--\ref{fig:dyn_results} corroborate the analytical results. The dynamic-cluster framework restores classical density-scaling sensing capacity in the small-cell-densified regime where the static framework plateaus.

\section{Conclusion}
In this paper, we developed a stochastic-geometry framework for cooperative ISAC target tracking under dynamic $K$-NN cluster re-selection, lifting the static-cluster assumption shared by all prior work. We proved that the antenna energy-conservation identity forces the mean BS-to-BS coupling gain to unity, identifying densification as an antenna-irreducible liability for monostatic sensing, and we derived a hard physical-layer beamwidth bound $\theta_m\!\le\!\theta_{\mathrm{crit}}(r_0)$ via a first-passage-time analysis of the single-BS Brownian-escape problem. By mapping the $K$-NN cluster re-selection onto Brownian motion with stochastic resetting, we obtained an exact closed-form expression $\mathbb{E}[\rv{T}_{\mathrm{loss}}^{\mathrm{dyn}}]\!=\!\nu_h^{-1}[I_0(R\sqrt{\nu_h/D})-1]$ for the cooperative mean tracking lifetime, which recovers the static percolation phase transition in the $\nu_h\!\to\!0$ limit and dissolves it under any positive handover rate. Combined with a per-link availability floor $K_{\mathrm{rel}}$, the dynamic framework restores classical linear density scaling of sensing capacity throughout the realistic 6G regime, while the static-cluster baseline plateaus at $\sim\!265\,\mathrm{targets/km^{2}}$ above $\lambda_b\!\approx\!80\,\mathrm{BS/km^{2}}$ --- a $\sim\!12\times$ capacity gap and $\sim\!13\times$ per-target cluster-commitment reduction at the small-cell densification ceiling $\lambda_b\!=\!10^{3}\,\mathrm{BS/km^{2}}$.

\enlargethispage{2\baselineskip}
\bibliographystyle{IEEEtran}
\bibliography{ref}

\end{document}